\documentclass{aa}  
\usepackage{graphicx}

\usepackage{txfonts}
\usepackage{natbib}
\usepackage{rotating}
\usepackage{array}
\usepackage{gensymb}
\usepackage{afterpage}
\bibpunct{(}{)}{;}{a}{}{,} % to follow the A&A style

\setcitestyle{authoryear,open={(},close={)}}
\usepackage{hyperref}
\usepackage{subcaption}

\newcommand{\kms}{km~s$^{-1}$}

\begin{document}

   \title{Gas inflow and outflow in an interacting high-redshift galaxy}

       \subtitle{The remarkable host environment of GRB 080810 at $z=3.35$}

   \author{P. Wiseman
	\inst{1}
	\and
	 D. A. Perley
	\inst{2}    
           \and
	 P. Schady
	\inst{1}
	\and
	J. X. Prochaska
	\inst{3}    
	\and
	A. de Ugarte Postigo
	\inst{4,5}
	\and
	T. Kr{\"{u}}hler
	\inst{1}
	\and
	R. M. Yates
	\inst{1}
	\and
	J. Greiner
	\inst{1}
          }

   \institute{Max-Planck-Institute f\"ur Extraterrestrische Physik (MPE), Giessenbachstrasse 1, 85748 Garching, Germany\\
	\email{wiseman@mpe.mpg.de}
	\and
	Astrophysics Research Institute, Liverpool John Moores University, IC2, Liverpool Science Park, 146            Brownlow Hill, Liverpool L3 5RF, UK
	\and
	 Department of Astronomy and Astrophysics, UCO/Lick Observatory, University of California, 1156 High Street, Santa Cruz, CA 95064, USA
	\and	
	Instituto de Astrofísica de Andalucía (IAA-CSIC), Glorieta de la Astronomía s/n, 18008 Granada, Spain
	\and
	Dark Cosmology Centre, Niels Bohr Institute, University of Copenhagen, Juliane Maries Vej 30, 2100 Copenhagen, Denmark
	}
  \date{Submitted: April 28, 2017; Accepted: July 13, 2017}
 \abstract{We reveal multiple components of an interacting galaxy system at $z\approx3.35$ through a detailed analysis of the exquisite high-resolution Keck/HIRES spectrum of the afterglow of a gamma-ray burst (GRB). Through Voigt-profile fitting of absorption lines from the Lyman-series, we constrain the neutral hydrogen column density to $N_{\ion{H}{I}} \leq 10^{18.35}$ cm$^{-2}$ for the densest of four distinct systems at the host redshift of GRB~080810, among the lowest $N_{\ion{H}{I}}$ ever observed in a GRB host, despite the line of sight passing within a projected 5 kpc of the galaxy centres. By detailed analysis of the corresponding metal absorption lines, we derive chemical, ionic and kinematic properties of the individual absorbing systems, and thus build a picture of the host as a whole. Striking differences between the systems imply that the line of sight passes through several phases of gas: the star-forming regions of the GRB host; enriched material in the form of a galactic outflow; the hot and ionised halo of a second, interacting galaxy falling towards the host at a line-of-sight velocity of 700 \kms; and a cool, metal-poor cloud which may represent one of the best candidates yet for the inflow of metal-poor gas from the intergalactic medium.}


   \maketitle
%
%________________________________________________________________

\section{Introduction}
Far from their historical description as `island Universes', galaxies interact continuously with their immediate surroundings and their broader environments, known as the circumgalactic medium (CGM) and intergalactic medium (IGM). Over the past 40 years it has been established that continued accretion of fresh, metal-poor gas is required to fuel ongoing star formation in galaxies over cosmological timescales \citep{Rees1977,White1978,White1991,Dekel2006}. This star formation can be quenched and regulated by gaseous outflows and fountains which enrich the IGM and CGM with metals, via regulatory winds powered by supernovae (SNe), stellar winds and active galactic nuclei (e.g. \citealt{Mathews1971,Larson1974,Martin1999,DallaVecchia2008,DallaVecchia2012}).
While observations of outflows have been made out to high redshifts \citep{Shapley2003,Martin2005,Weiner2009,Rubin2014,Finley2017}, it is much harder to detect emission from infalling streams due to their lower surface brightness as a result of lower metallicity and volume filling factor, although it has been possible to identify infalling gas through absorption lines in individual galaxy spectra \citep{Rubin2012}.

Rather than observe galaxies and their CGM by their emission, it is possible to measure their properties by absorption if they line up with a bright, background point source such as a quasar (QSO) or gamma-ray burst (GRB). At sufficiently high redshift\footnote{Due to the large optical depth of the Earth's atmosphere to wavelengths bluer than $\sim 3000~\AA$, Ly-$\alpha$ is only visible from the ground at $z\gtrsim 1.6$}, some quasar spectra show the imprint of a dense column of neutral hydrogen in the form of a damped Lyman-alpha absorber (DLA), defined as a system with neutral hydrogen column density $\log N_{\ion{H}{I}} >20.3$ \citep{Wolfe2005}\footnote{Throughout the paper, column densities $(N)$ are given in units of cm$^{-2}$}.

 DLAs are associated with intervening galaxies between the observer and the QSO \citep{Wolfe1986,Prochaska1998}, and are one of the most prominent tools used to study the properties of cold gas at high redshift \citep{Dessauges-Zavadsky2006,Rafelski2012,Rafelski2014,DeCia2016}. GRBs are produced during the deaths of very massive stars \citep{Woosley2006a}, as proven by their association with type Ic supernovae (SNe; \citealt{Galama1998,Hjorth2003,Greiner2015,Cano2017}). While QSO-DLAs tend to probe dense pockets in the outskirts of galaxies, the DLAs commonly seen in the afterglow spectra of GRBs represent the dense ISM and star-forming regions of their host galaxies \citep{Fynbo2006,Prochaska2007b,Ledoux2009,Kruehler2013}. Along with detections of metal absorption lines redward of the Lyman-alpha (Ly-$\alpha$) feature, these absorption spectra can be used to calculate the metallicity (e.g. \citealt{Savaglio2012,Cucchiara2015,Wiseman2017a}) and study the kinematics of the absorbing material (e.g. \citealt{Prochaska2008a}). %, and are found to have systematically lower metallicity than GRB-DLAs \citep{Prochaska2007c,Fynbo2013,Cucchiara2015}.

Because QSO lines of sight (LOS) are random, there is no preference for the occurrence of DLAs. There are thus large samples of QSO-sub-DLAs, which are also known as super Lyman limit systems (SLLS; $19<\log N_{\ion{H}{I}}<20.3$; e.g. \citealt{Peroux2007,Quiret2016}). Similar studies extend to Lyman limit systems (LLS; $17.3<\log N_{\ion{H}{I}}<19$; e.g.  \citealt{Prochaska2015,Fumagalli2016,Lehner2016}), and partial Lyman limit systems (pLLS) and the Ly-$\alpha$ forest ($\log N_{\ion{H}{I}}<17.3$; e.g. \citealt{Aguirre2003,Pieri2013}), all of which tend to be representative of different phases of the ISM, CGM and IGM respectively. LLSs have been the subject of particular interest recently as they are likely to represent absorption through dense, cool pockets of the predominantly hot and ionised CGM. Simulations and observations suggest that they may trace the fresh gas that fuels star formation \citep{Fumagalli2011a,Fumagalli2011b,Glidden2016,Lehner2016a}, as well as outflowing, metal-rich gas \citep{Lehner2013}, or indeed a combination of both \citep{Fumagalli2013}. 

\begin{figure*}
   
   %\centering
   \includegraphics[width=19cm]{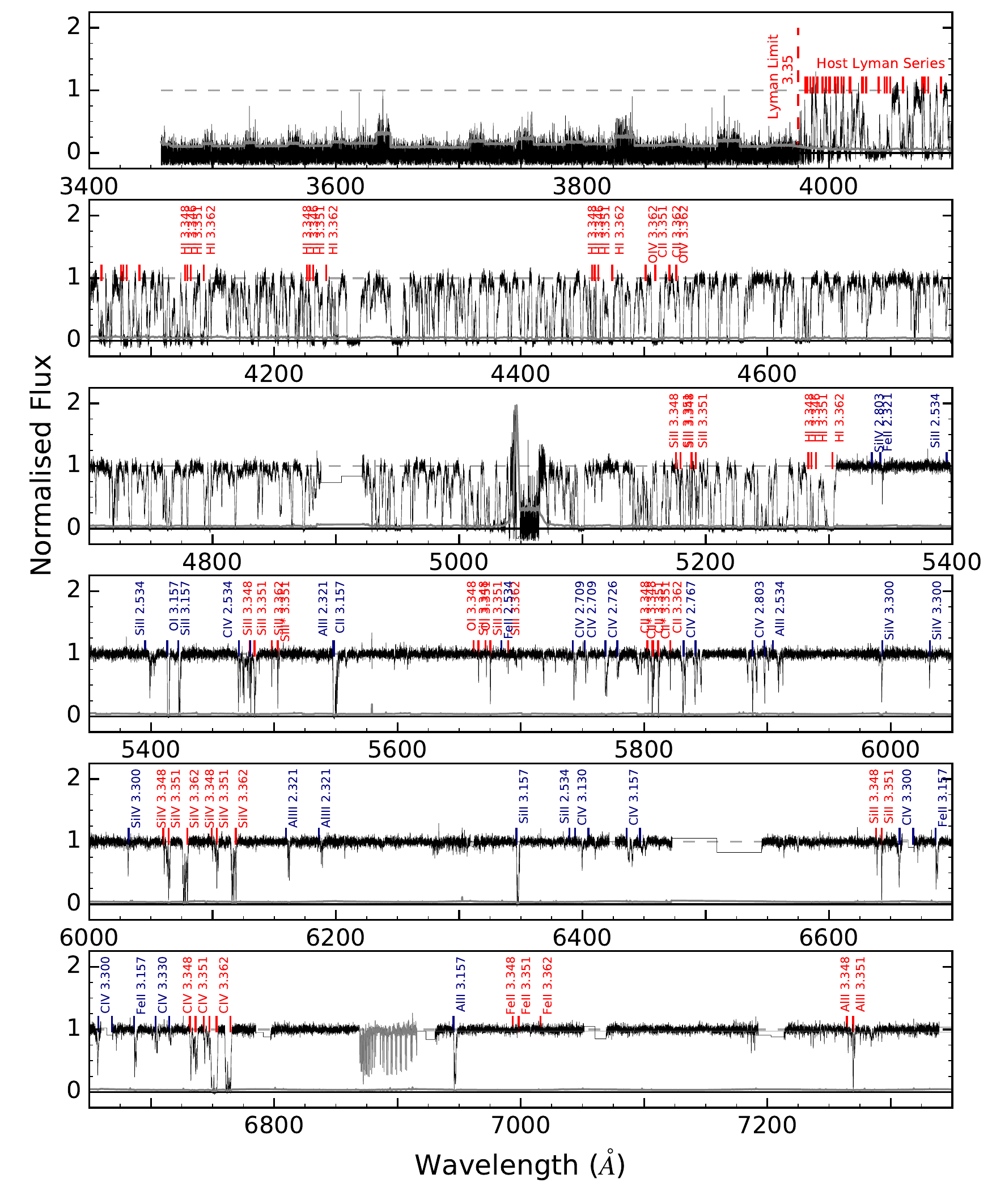}
   
   \caption{\label{fig:spec_full} \small The full HIRES spectrum of GRB 080810. Absorption from the host system complex around $z=3.35$ is marked in red, while a selection of the strongest intervening systems have been marked in blue colours. For clarity, not all Ly-$\alpha$ forest lines are marked. Absorption corresponding to telluric lines has been coloured grey. The Lyman limit is clearly visible just short of 4000 $\AA$, as is the termination of Ly-$\alpha$ absorption at 5300 $\AA$.}
\end{figure*}
%Since the turn of the century, rapid-response observations of long gamma-ray Bursts (GRBs) and their afterglows has facilitated deep insights into the distant Universe.  which can be observed out to very high redshift \citep{Tanvir2009}. Extensive follow up programmes has allowed the statistical study of GRB host galaxies and their role as a probe of star formation throughout the Universe \citep{Hjorth2012,Schulze2015,Kruehler2015,Perley2016a,Perley2016b,Vergani2017} Complimentary to the targeted follow up of host galaxies, the featureless synchrotron spectra of GRB afterglows \citep{Meszaros1997} also allows the study of absorbing and extincting systems along the line of sight. The subject of these studies ranges from the interstellar medium (ISM) of the GRB host galaxy (e.g. \citealt{Prochaska2007,Kruehler2013,DElia2014} to that of the Milky Way \citep{Watson2011}, and can focus on properties such as gas and metal content \citep{Ledoux2009,Cucchiara2015} and dust characteristics \citep{Savaglio2004,Greiner2011,Zafar2013,DeCia2013,Wiseman2017}.

While the physical conditions of GRB-DLAs and sub-DLAs have been extensively studied (e.g. \citealt{Prochaska2007a,DElia2011,Schady2011,Sparre2014,Friis2015,Wiseman2017a}), there is much less known about the systems hosting GRBs with a much lower hydrogen column density. Typically, they are left out of studies based on metals and dust due to the complicated effects of ionisation, which at higher column densities is rendered negligible \citep{Viegas1995,Wolfe2005,Peroux2007}. They are also much rarer. Of the 75 GRBs identified by \citet{Cucchiara2015} as having measurable $\ion{H}{I}$ and at least one metal line, only 5 ($6.7\%$) qualify as LLSs. Of these, \object{GRB 060124} ($\log N_{\ion{H}{I}}=18.5$; \citealt{Fynbo2009}) and \object{GRB 060605} ($\log N_{\ion{H}{I}}=18.9$; \citealt{Ferrero2009}) were observed with only low-resolution spectrographs and provide little information regarding their hosts. \object{GRB 060607A} ($\log N_{\ion{H}{I}}=16.8$) and \object{GRB 080310} ($\log N_{\ion{H}{I}}=18.8$) have high-resolution spectra from VLT/UVES, and have been studied in \citet{Fox2008}, focusing on high-ion absorption, particularly the prominence of blue-shifted absorption tails of those species. 

In this paper, we present a detailed study of \object{GRB 080810}, whose spectrum shows a very small total $N_{\ion{H}{I}}$ distributed along several absorption components at the host redshift which show dramatically different chemical and physical properties. Throughout the manuscript we use solar abundances from \citet{Asplund2009}. We assume a flat $\Lambda$CDM cosmology with \emph{Planck} parameters: $H_0 = 67.3$ km s$^{-1}$ Mpc$^{-1}$, $\Omega_{\mathrm{m}}=0.315$, and $\Omega_{\Lambda}=0.685$ \citep{PlanckCollaboration2014}. Errors are given at the 1 $\sigma$ confidence level.

%__________________________________________________________________

\section{Observations and Data Reduction\label{sec:obs}}
On 2008-08-10 at $T_0$=13:10:12 UT, the Burst Alert Telescope (BAT; \citealt{Barthelmy2005}) on board \textit{Swift} \citep{Gehrels2004} triggered on GRB 080810 \citep{Page2008}, which at $T_0 + 80$s was detected as a bright source in \textit{Swift}'s X-ray Telescope (XRT; \citealt{Burrows2005}) and Ultra-violet and Optical Telescope (UVOT; \citealt{Roming2005}). The burst was localised to RA = 23:47:10.48, Dec. =+00:19:11.3 (J2000+/- 0.6"), and \citet{Page2009} provide an overview of the prompt and subsequent follow-up observations carried out by numerous ground- and space-based observatories, and we refer the reader to that paper for a detailed analysis of the broadband spectral evolution of the GRB. 

\subsection{Keck/HIRES afterglow spectrum\label{subsec:obs_keck}}

Starting 37.6 minutes after the trigger at 13:47:50 UT, GRB 080810 was observed with the High Resolution Echelle Spectrometer (HIRES; \citealt{Vogt1994}) mounted on the 10-metre Keck I telescope of the W. M Keck Observatory located at the summit of Mauna Kea, Hawaii. A series of two exposures of 1000\,s each were taken using the C5 decker, providing a FWHM spectral resolution of $\approx 8$ \kms. The data were reduced with the HIRedux data reduction pipeline bundled within the XIDL software package\footnote{https://github.com/profxj/xidl}. Full details on the data reduction algorithms are given in \citet{Bernstein2015}. The data were coadded optimally and normalized with low-order polynomial fits to individual echelle orders. The extracted 1D spectrum is shown in Fig. \ref{fig:spec_full}.

\subsection{Late time imaging and spectral observations\label{subsec:obs_img}}

We observed the host galaxy of GRB 080810 on several occasions. On 2014-07-22 we acquired $6\times 250$\,s of imaging in the $r$-band with the Optical System for Imaging and low-Intermediate-Resolution Integrated Spectroscopy (OSIRIS; \citealt{Cepa2000}) on the 10.4 m Gran Telescopio Canarias (GTC). We 
acquired imaging with the Low-Resolution Imaging Spectrograph (LRIS; 
\citealt{Oke+1995}) mounted at Keck, in the $B$, $G$, $R$, and $RG$\,850 (similar to SDSS-$z$) 
filters on the nights of 2014-08-30 and 2014-08-31 (three exposures in 
each filter totaling between six and nine minutes total on-source).  A 
longslit spectrum was also acquired at a position angle of 35 degrees, 
with a total exposure time of 2400\,s in the blue arm and 2200\,s in the 
red arm.    We also acquired near-infrared imaging using the 
Multi-Object Spectrograph For Infrared Exploration (MOSFIRE; 
\citealt{McLean+2012}) on 2014-10-01 using the same telescope, in the 
$K_s$ and $J$ bands.   In addition, on 2015-07-23 we acquired a further 600\,s of imaging with GTC/OSIRIS
 in the $i$-band, and we obtained deep FOcal Reducer/low dispersion Spectrograph 2 (FORS2) $R$-band imaging from the VLT archive (originally published in \citealt{Greiner2015a}). Finally, we have obtained \emph{Spitzer} 3.6 $\mu$m data as part of the SHOALS survey \citep{Perley2016b}.

%__________________________________________________________________

\section{Analysis of the afterglow spectrum \label{sec:analysis}}
The normalised spectrum of GRB 080810 (Fig. \ref{fig:spec_full}) shows negligible flux bluewards of a cut-off at 3795 $\AA$, corresponding to the Lyman-limit at $z\approx3.35$. Redwards of this is a dense Ly-$\alpha$ forest, which terminates with two deep Ly-$\alpha$ absorption troughs around 5300 $\AA$, assumed in \citet{Page2009} to correspond to the GRB host system at an approximate redshift of 3.35. Further redwards, the spectrum is populated by mild metal absorption from the host systems, as well as various metal lines associated with lower redshift systems. 

\subsection{Defining the host complex at $z=3.35$\label{subsec:host_lines}}

The large column densities in GRB-DLAs and sub-DLAs cause, through quantum mechanical effects, the Ly-$\alpha$ absorption feature to be saturated and often to span regions over 15 $\AA$ wide about its central rest-frame wavelength of 1215.8 $\AA$. A fit to the damping of the red wing is then used to measure the $N_{\ion{H}{I}}$ for the entire host. The deep, wide trough from the DLA means that any further Ly-$\alpha$ absorption only becomes evident at much shorter wavelengths, shifted significantly in redshift space (and thus much closer to Earth) or corresponding to a shift of several thousands of \kms~in velocity space at the redshift of the DLA. This blueshifted Ly-$\alpha$ absorption is thus totally unrelated to the GRB or its host galaxy and surrounding environment.
The spectrum of \object{GRB 080810} shows no such DLA, and therefore there is no clear distinction between the absorption components which are related to the host and those which are Ly-$\alpha$ forest absorption from the IGM. 

%One option is to impose a simple redshift cut-off. Here, the assumption is that the systems above this cut are spatially close, their absorption lines shifted purely by local peculiar velocities. A typical, rich cluster of galaxies has a velocity dispersion of $\sigma\approx 1000$ \kms. Although this system is far from being a cluster, we can use this as a rough limit, above which fall four systems:  

\begin{figure*}
   
   	 {\includegraphics[width=19cm]{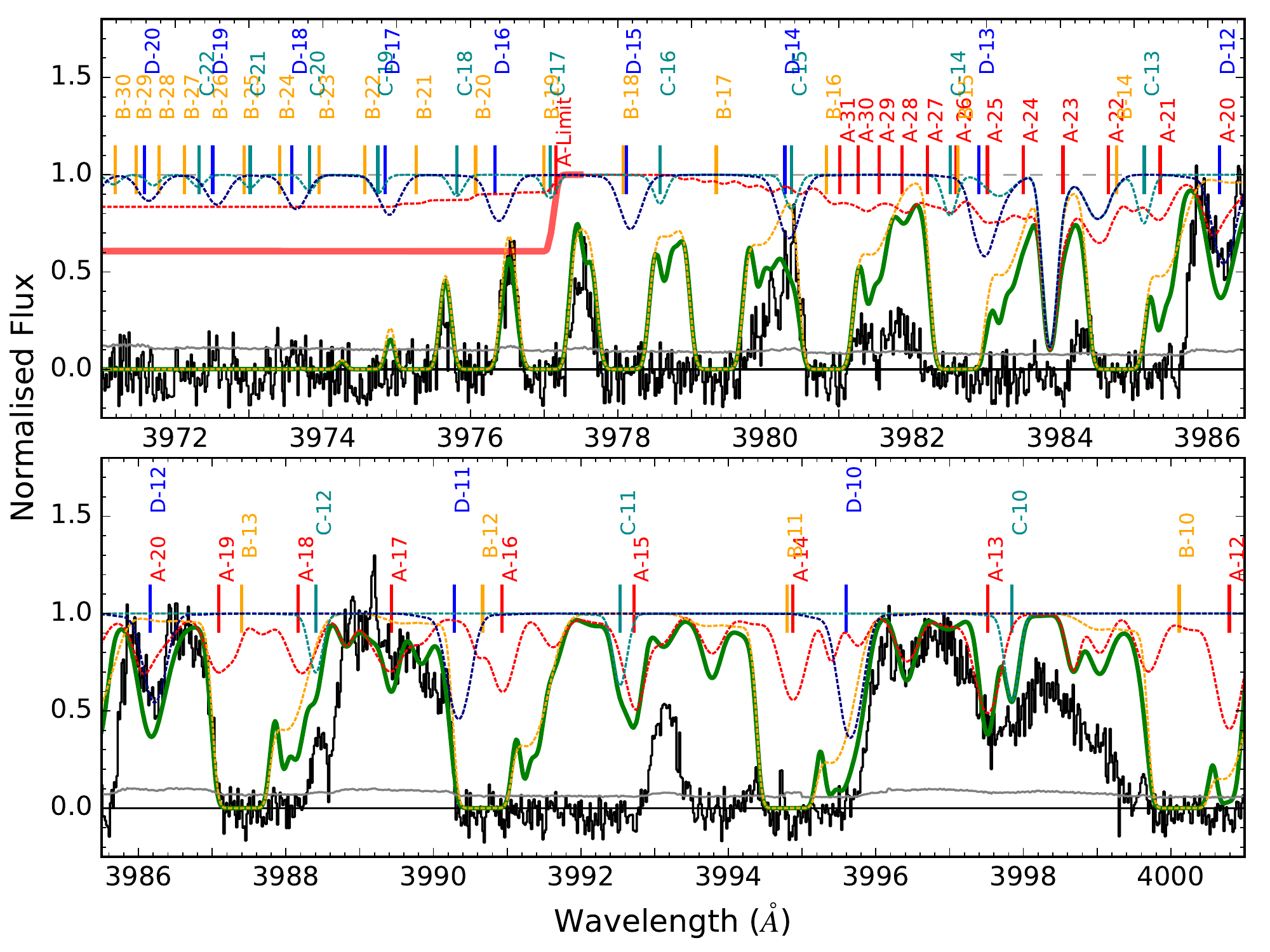}
   
      \caption{\small The normalised HIRES spectrum of \object{GRB 080810} (black) around the Lyman limit of 911.8 $\AA$. Fits to the absorption by $\ion{H}{I}$ by the four systems of the host are plotted by dashed lines in red, orange, cyan, and dark blue for systems A, B, C, and D respectively, with the combined fit shown in solid green. Labels show the location of each order for each component. The error spectrum is shown in grey. The thick red line corresponds to the upper limit in system A of $N_{\ion{H}{I}}\leq16.9$.}
	\label{fig:ly_limit}}
 \end{figure*}

 \begin{figure}
   
 \centering
  \includegraphics[width=8.5cm]{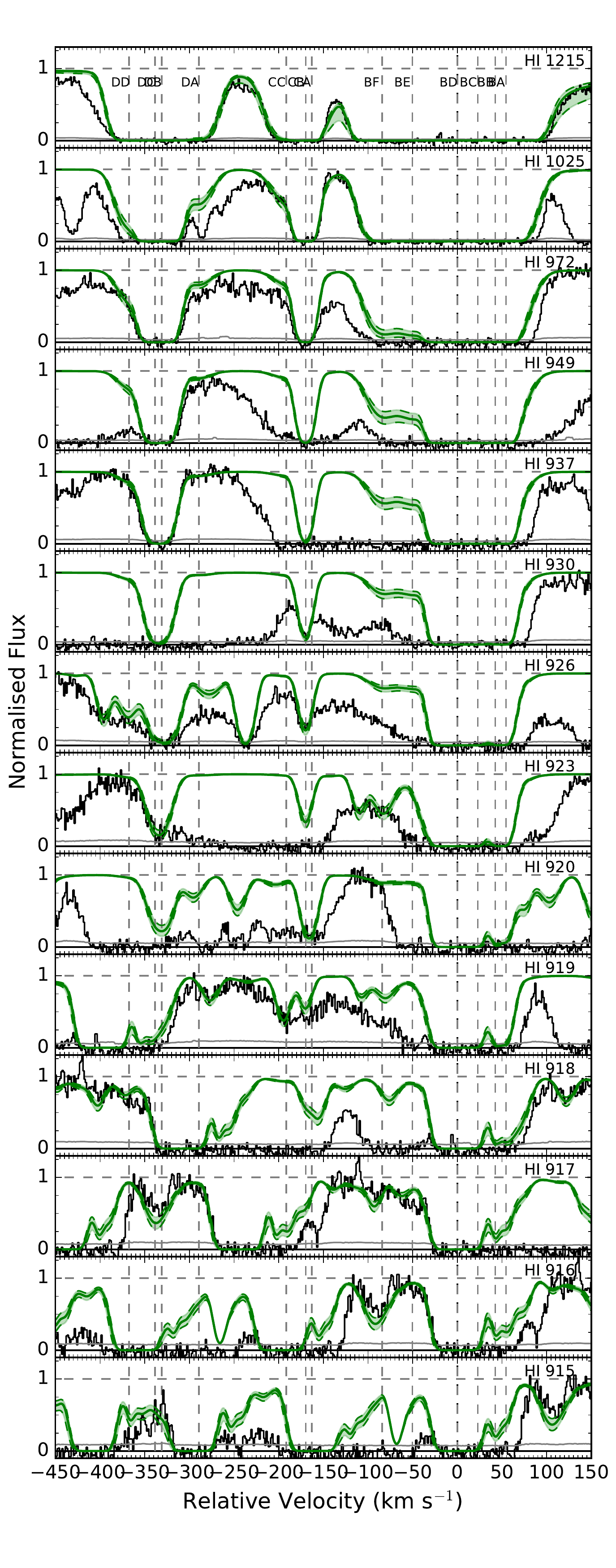}
   
   \caption{\label{fig:lyman_BCD} \small A selection of fits to the Lyman series for systems D, C, and B at $z=3.346$, 3.348 and 3.351 respectively. Lines between the third and tenth orders allow systems C and D to be constrained well. Higher order Lyman-series transitions near the Lyman limit, where system B better is constrained, are shown in Fig. \ref{fig:ly_limit}. Shaded regions show 0.3 dex deviations from the best fit. Grey vertical lines represent the centres of the corresponding components.}
   \end{figure}

\begin{figure}
   \centering
   \includegraphics[width=8.5cm]{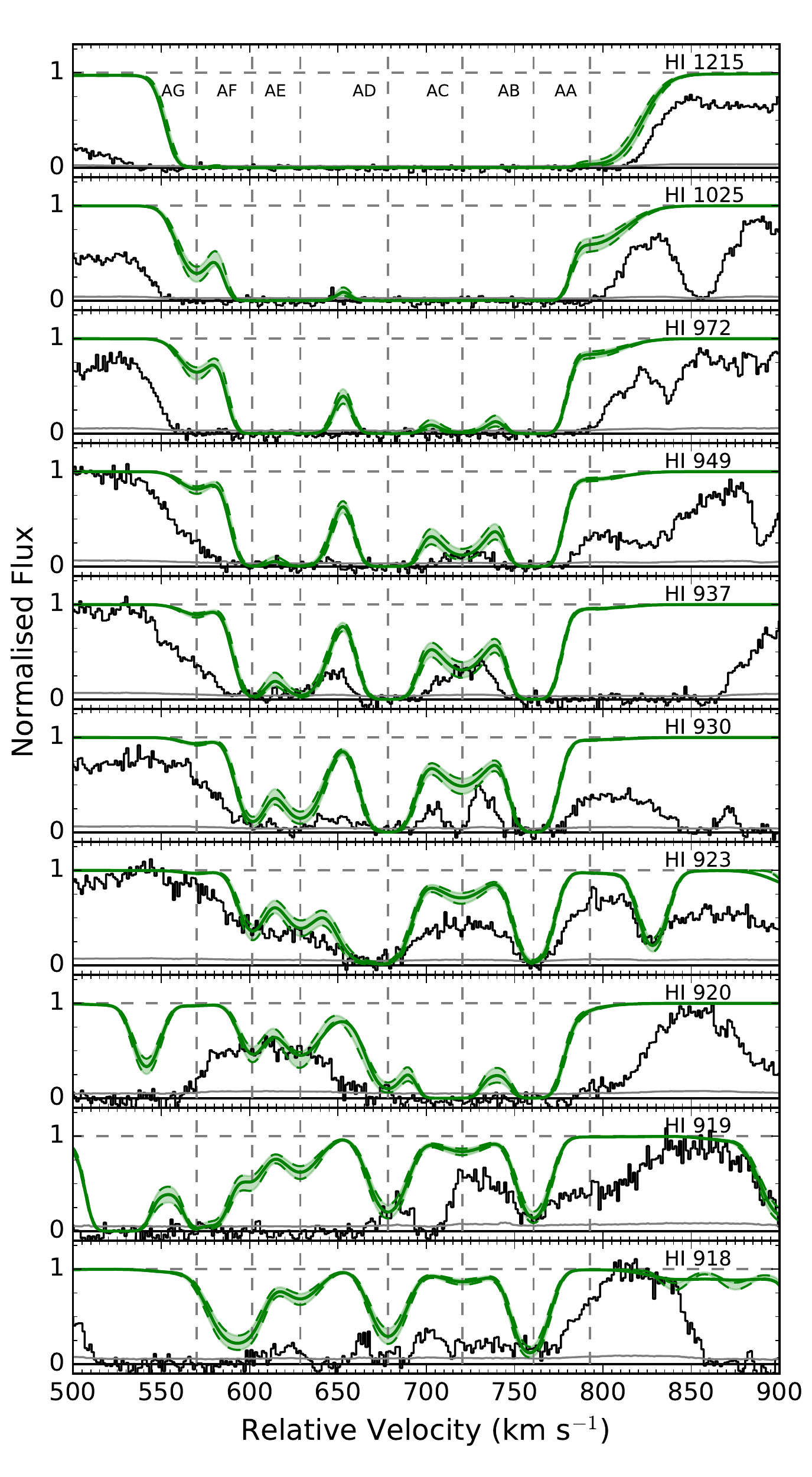}
   
  \caption{\label{fig:lyman_A}\small  A selection of fits to the Lyman series for component A at $z=3.36$. $\ion{H}{I}$ 937, 930, 923, and 920 are used to constrain components AD-AG, while the transitions at 923, 919 and 918 $\AA$ provide constraints on the main red component AB. Shaded regions show 0.3 dex deviations from the best fit. Grey vertical lines represent the centres of the corresponding components.} 
   \end{figure}

The reddest Ly-$\alpha$ absorption is centered around $z=3.361$, which we denote system A. Roughly 700 \kms~bluer at $z=3.351$  lies system B. These two objects are also detected in the late-time imaging and low-resolution spectroscopy, as detailed in Sect. \ref{subsec:imaging}. Furthermore, based on our subsequent analysis (Sect. \ref{subsec:discuss_bc}), there is strong evidence to show that system B is in fact the host of the GRB. This means that A is in the foreground, and implies a positive relative velocity between A and B of $\approx 700$ \kms. To simplify the analysis, we henceforth denote B as being at rest in relation to the Hubble Flow, and thus define it as the point where peculiar velocity $v=0$ \kms.

Further bluewards, systems C and D are centred at $z_{\textnormal{\textsc{c}}}=3.348$ and $z_{\textnormal{\textsc{d}}}=3.346$, a shift of $-170$ \kms~and $-340$ \kms~respectively with respect to B. These velocities satisfy the (observationally motivated) condition used by \citet{Prochaska2015} to define an LLS, in that they lie within 500 \kms~of the point of deepest absorption and are thus treated as part of the same system. However, if we assume that C and D have no local velocity relative to B, they are 0.6 Mpc and 1 Mpc displaced from B respectively, while if instead we take A and B to be systems falling at equal and opposite peculiar velocities towards each other, the respective separation to C and D increases to 1.6 and 2 Mpc. These values lie significantly outside the average virial radius of 105 kpc for a dark matter halo containing a galaxy of the mass $\log M_* =10^{10} M_{\odot}$ \citep{Henriques2015}, a similar stellar mass to that which we derive for this host in Sect. \ref{subsec:imaging}. Since we are unable to break the degeneracy between local velocities and cosmological redshift, it is not possible to determine for certain, based on redshift arguments alone, whether these two systems are also bound to the A-B system, but include them in our analysis in an attempt to determine their nature. We return to this topic in the discussion section, where we use chemical and physical arguments to locate the systems in space. Lyman-series absorption from the components A-D is shown in Figs. \ref{fig:lyman_BCD} and \ref{fig:lyman_A}. 

The next absorption systems, E and F, are found at $z_{\textnormal{\textsc{e}}}=3.338$ and $z_{\textnormal{\textsc{f}}}=3.330$ respectively. Assuming they are bound to the host system, they have relative velocities of  $\approx-1000$ and $-1550$ \kms  ~with respect to B, and thus fall outside the aforementioned criterion to be assumed as part of the same LLS. Assuming on the other hand that they are static relative to the expansion of the Universe, the separation between E (F) and the mean redshift of A and B is $\sim$3.6 Mpc ($\sim$5.2 Mpc). Finally, there is no apparent low- or high-ionisation metal absorption associated with these redshifts, reducing the possibility that they are high velocity outflows from the GRB progenitor or the host galaxy system. We conclude that E and F are very unlikely to be part of the host system and are probably located in the IGM. We therefore define the host complex as being comprised of systems A, B, C and D.

\begin{figure}
   
   \centering
   \includegraphics[width=8.5cm]{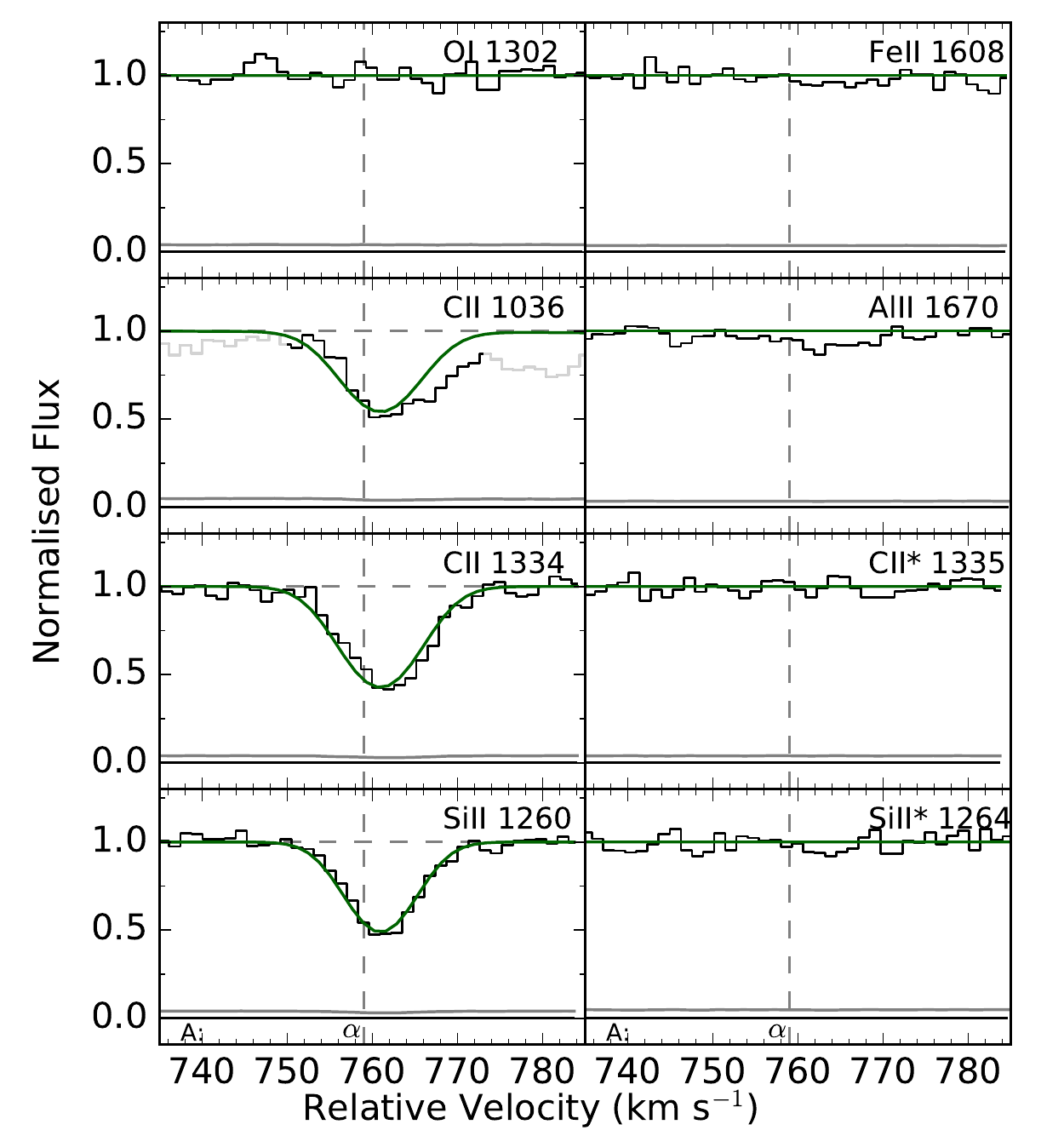}
   
   \caption{\label{fig:loion_A} \small The low-ionisation metal transitions in system A with their respective Voigt-profile fits in green. Light grey data represents unrelated absorption, while the error spectrum is plotted in dark grey. Vertical dashed lines correspond to the respective velocity components. This scheme is used in Figs. 5-11.} 
   \end{figure}
 \begin{figure}
   
   \centering
   \includegraphics[width=8.5cm]{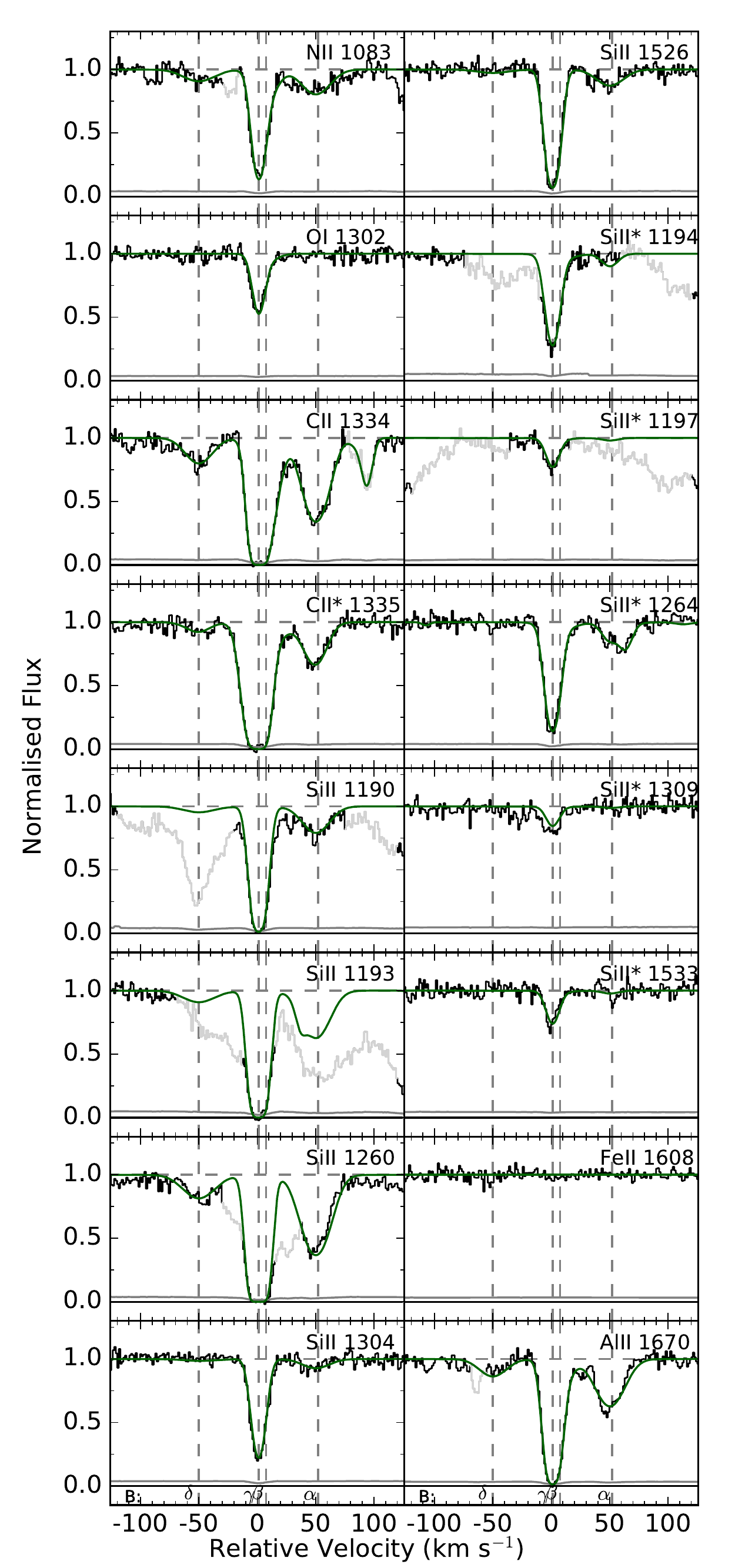}
   
   \caption{\label{fig:loion_B} \small The low-ionisation metal transitions in system B with their respective Voigt-profile fits in green. }
   \end{figure}
\begin{figure}
   
   \centering
   \includegraphics[width=8.5cm]{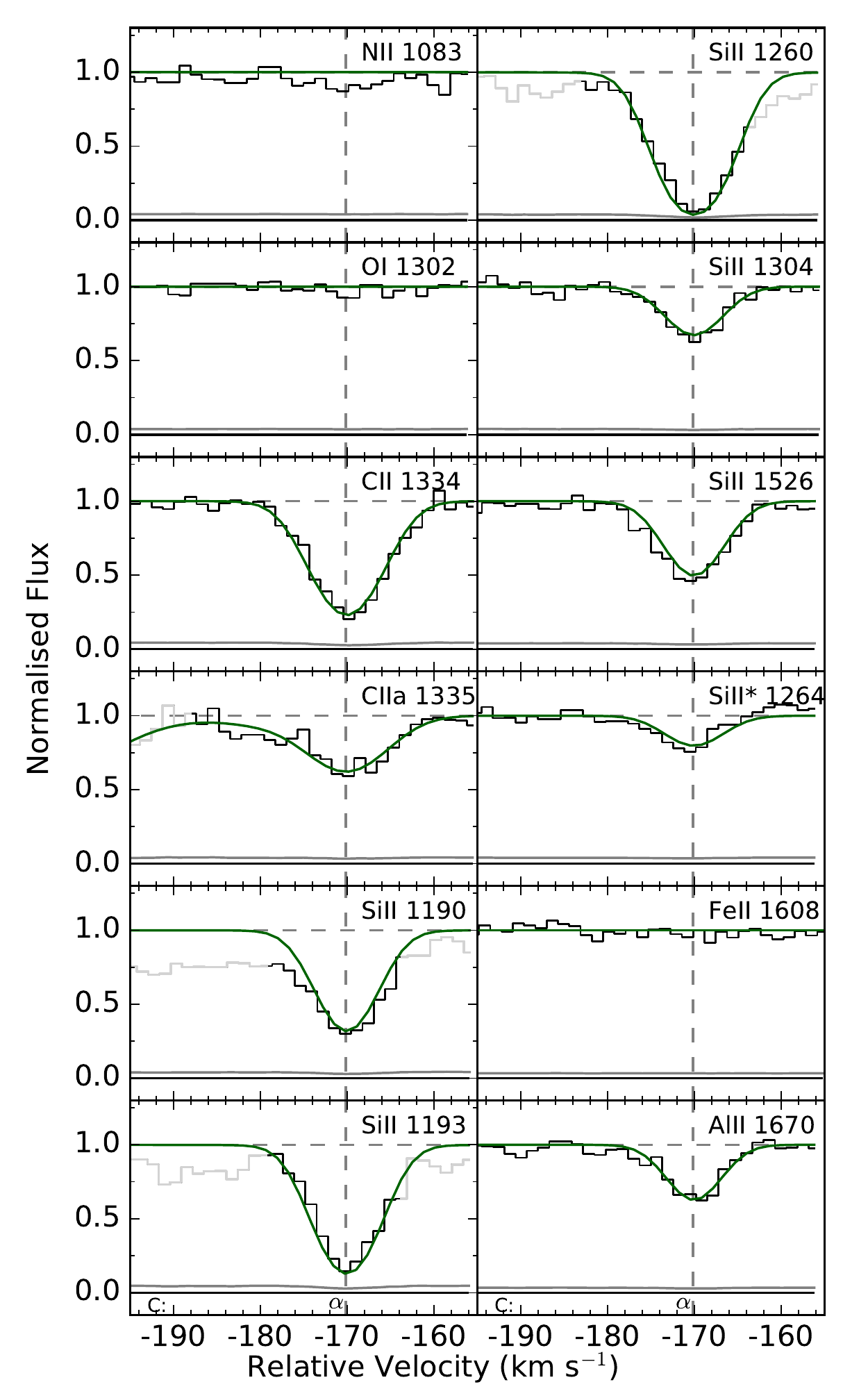}
   
   \caption{\label{fig:loion_C}\small  The low-ionisation metal transitions in system C with their respective Voigt-profile fits in green. }
   \end{figure}

\begin{figure}
   
   \centering
   \includegraphics[width=8.5cm]{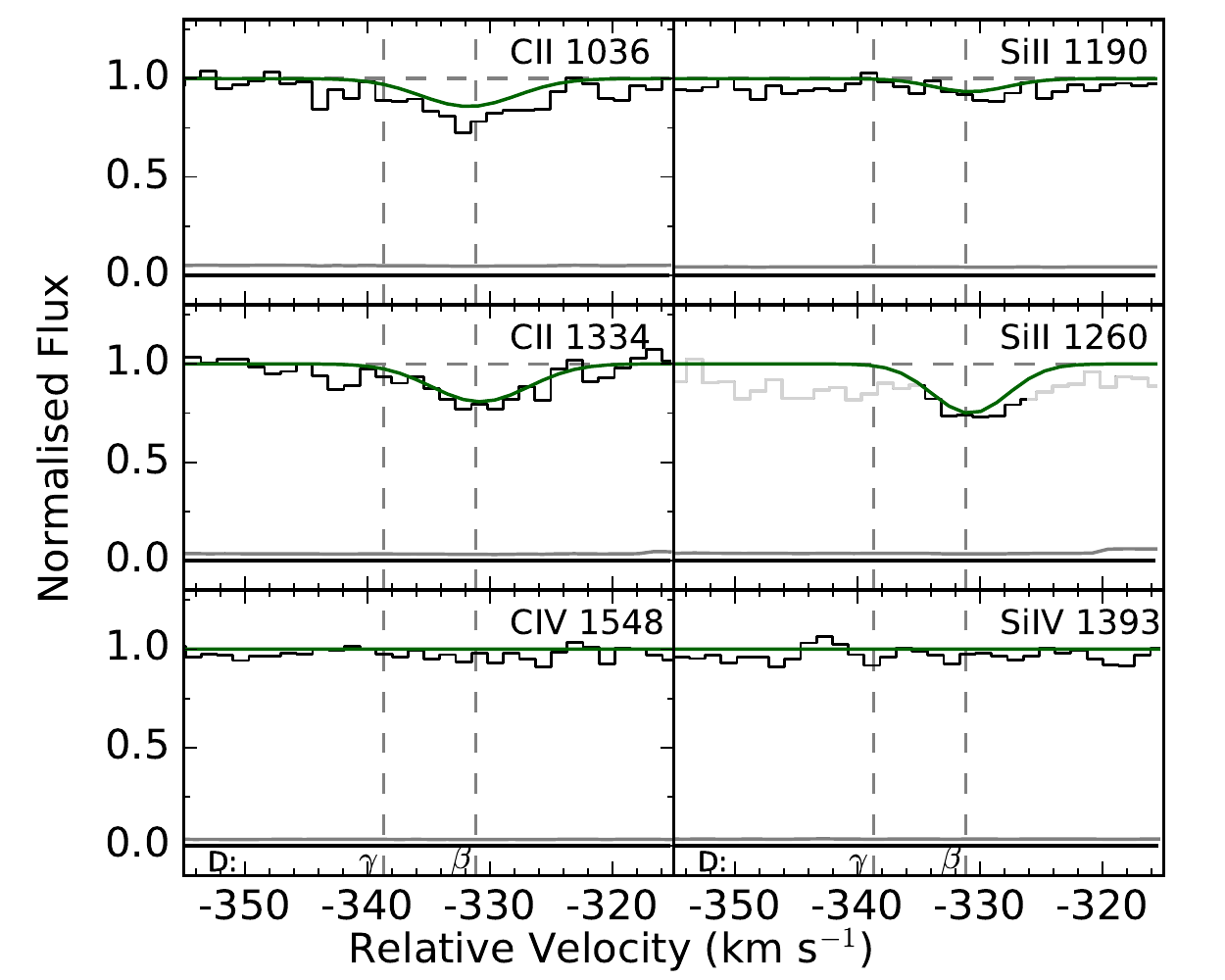}
   
   \caption{\label{fig:lonhi_D} \small The marginal detection of $\ion{C}{II}$ is shown along side other metal transitions which provide limits in system D. }
   \end{figure}

\begin{figure}
   
   \centering
   \includegraphics[width=8.5cm]{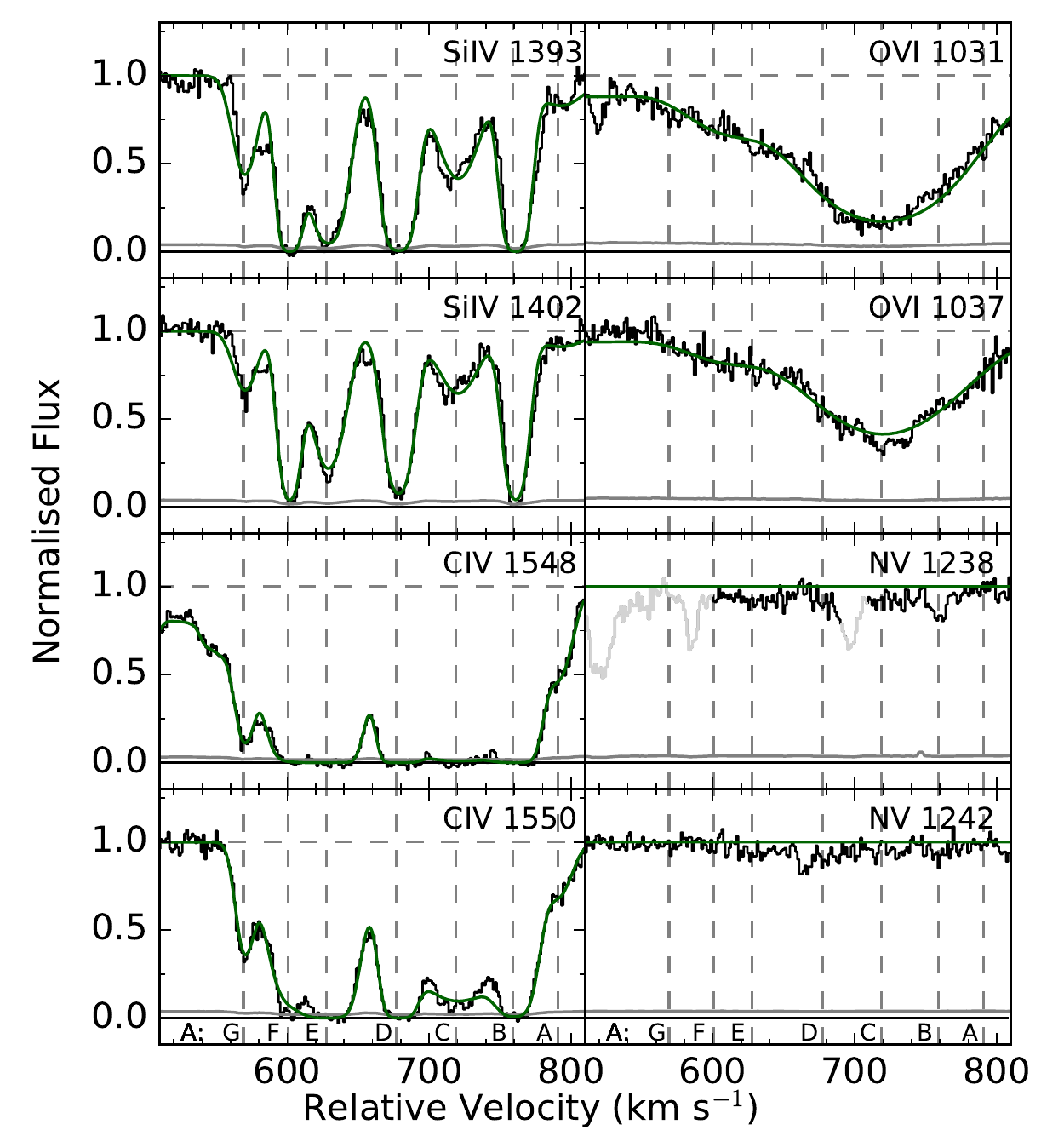}
   
   \caption{\label{fig:hiion_A} \small The high-ionisation metal transitions in system A with their respective Voigt-profile fits in green. }
   \end{figure}

\begin{figure}
   
   \centering
   \includegraphics[width=8.5cm]{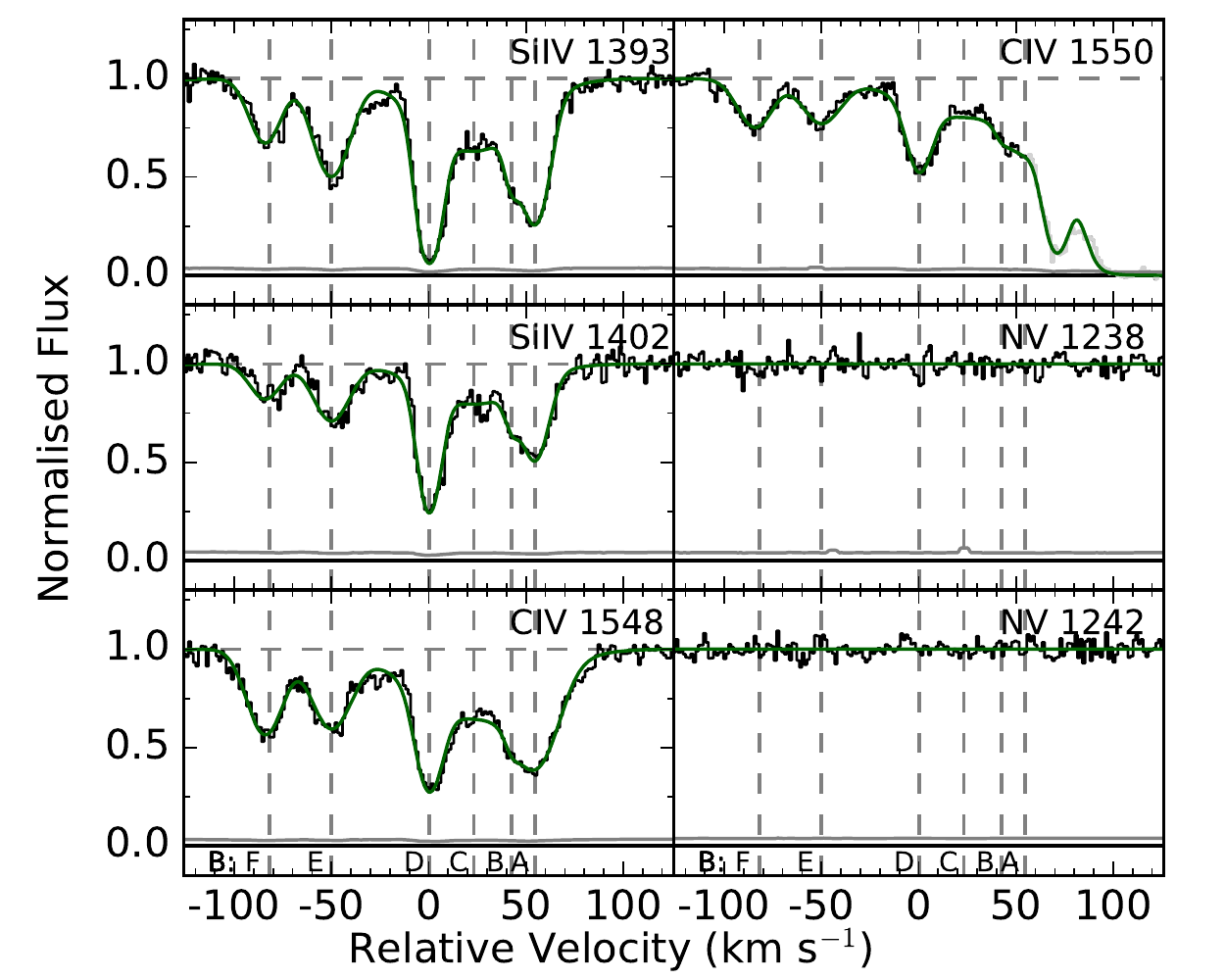}
   
   \caption{\label{fig:hiion_B} \small The high-ionisation metal transitions in system B with their respective Voigt-profile fits in green. }
   \end{figure}

\begin{figure}
   
   \centering
   \includegraphics[width=8.5cm]{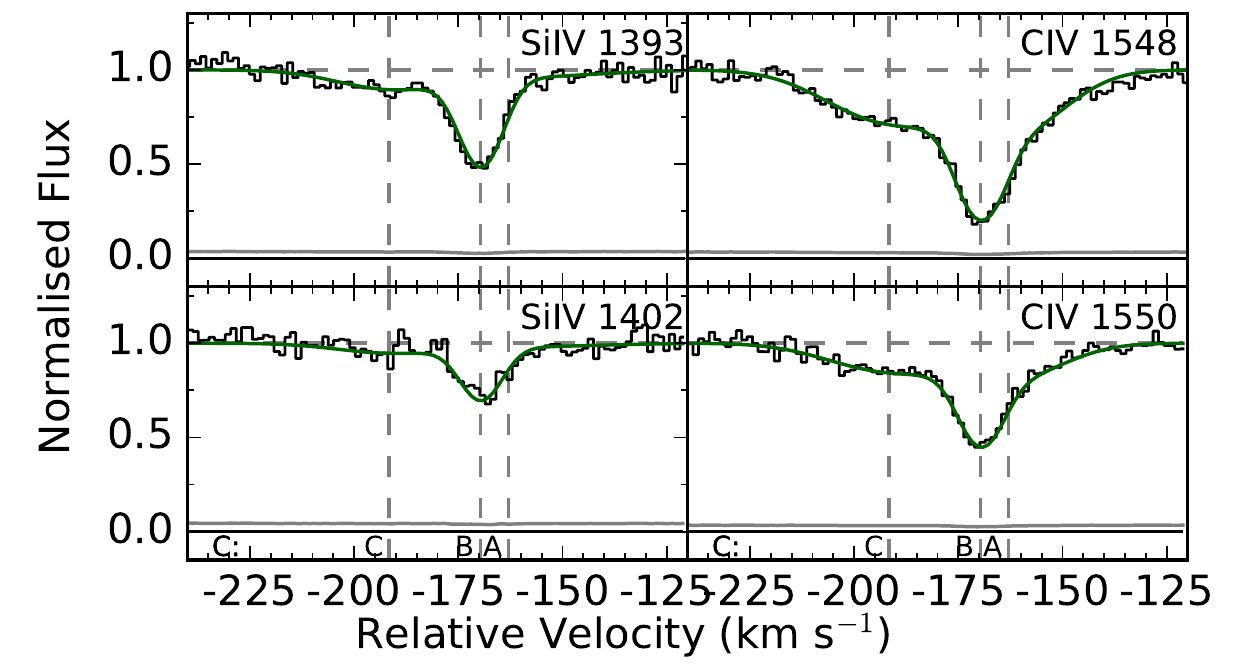}
   
   \caption{\label{fig:hiion_C} \small  The high-ionisation metal transitions in system C with their respective Voigt-profile fits in green. }
   \end{figure}

\subsubsection{$\ion{H}{I}$ column densities\label{subsubsec:HI}}
Unusually for a GRB host galaxy, there is no damping of the Ly-$\alpha$ lines in the presumed host of GRB 080810, which places an initial upper limit of $\log N_{\ion{H}{I}}< 19 $ for each of the systems. As was the case in \citet{Page2009}, a lower limit of $\log N_{\ion{H}{I}} > 17.3$ may be imposed due to 100\% absorption below the Lyman limit. With Ly-$\alpha$ being saturated but not damped, a Voigt-profile fit to this line alone does not constrain $N_{\ion{H}{I}}$ much further. However, due to the excellent signal-to-noise ratio (S/N) and high resolution of the spectrum, we are able to use higher order Lyman series lines to place much tighter constraints on the $ N_{\ion{H}{I}}$. A selection of these transitions is shown in Fig \ref{fig:lyman_BCD} for system B, C and D, and in Fig. \ref{fig:lyman_A} for system A. A more detailed view of the highest order Lyman series absorption and the Lyman limit is presented in Fig. \ref{fig:ly_limit}. 

The flux drops to zero around 3975 $\AA$, which is consistent with the Lyman limit at a redshift of 3.3598. However, the higher orders of the Lyman series of system A are unsaturated, and there is also non-zero flux blueward of 3977 $\AA$, which corresponds to the Lyman Limit of the strongest metal absorber in system A at $z=3.36206$. This implies that system A is a pLLS. The thick red line in in Fig. \ref{fig:ly_limit} corresponds to the flux drop caused by a column density of $\log N_{\ion{H}{I}}=16.9$ at the redshift of system A, thus placing this value as a strict upper limit. We deduce that the zero-flux at 3975 $\AA$ is due to the saturated Ly-21 and 22 transitions from system B, which has a significantly larger column density.

Each of the four host systems are split up further into components at slightly different redshifts and with different column densities and other physical parameters. In order to measure column densities more accurately, it is necessary to define these velocity components. Usually, absorption lines from species with similar ionisation potentials show very similar profiles, defined by their redshift $z$ and broadening parameter $b$. The only remaining free parameter is then $N$, which can be varied to find the best fit for all of the available lines of the species in question. However, determining $z$ and $N$ for $\ion{H}{I}$ is non-trivial. $\ion{H}{I}$ lines are strong and saturate up to the highest orders even for low column densities. Because it is so abundant, the stronger $\ion{H}{I}$ lines (Ly-$\alpha$, $\beta$, $\gamma$) often show absorption at large velocities from their central redshift, where there is no corresponding metal absorption. Additionally, the Lyman series is strongly contaminated by the Ly-$\alpha$ forest, so there is a degree of uncertainty created by not knowing whether absorption is coming from the host or from intervening systems.

While $\ion{H}{I}$ is expected to follow components seen in the neutral and singly-ionised metals (which are labelled with Greek letters), this is clearly not the case in system A: $\ion{C}{II}$ and $\ion{Si}{II}$ are only detected in one component, A$\alpha$, while $\ion{H}{I}$ covers over 250\, \kms~in velocity space. The typical $b$-parameter of absorption systems in GRB hosts is $\sim 12$~\kms  \citep{DeUgartePostigo2012}, so $\ion{H}{I}$ in A is clearly formed of several components. Due to its low column density relative to most GRBs, many of the transitions from $\ion{H}{I}$ 949 to the Ly-limit show sections of non-zero flux, so some velocity structure can be determined. It is apparent in Fig. \ref{fig:lyman_A} that the deeper absorption is matched to the central redshifts of the high-ionisation metals (high-ions; Sect. \ref{subsec:metals}), whose absorption components spread across the entire velocity range traced by $\ion{H}{I}$.

Proceeding with caution, we use the Voigt-profile fitting software \texttt{VPFit (v.10.2)} \footnote{VPFit, R. F. Carswell \& J. K. Webb, 2014: http://www.ast.cam.ac.uk/ rfc/vpfit.htm}, with initial guesses for absorption lines at redshifts and $b$ parameters taken from the high-ions, and $N$ given by the upper limit measured from the Ly-limit. The fits to the entire available Lyman series (some lines such as  $\ion{H}{I}$ 926 are totally saturated and thus not used, since the amount of contamination from one or several Ly-$\alpha$ forest lines is unknown) are then simultaneously adjusted by eye until a satisfactory fit is found (this technique is commonly used for fitting of $\ion{H}{I}$, e.g. \citealt{Prochter2010,Rafelski2012,Jorgenson2013,Prochaska2015}). This is necessary because of Lyman-alpha forest contamination, which is impossible to account for in the fitting routine. The simultaneous fit is possible as the components are relatively narrow and thus independent of each other. The best fit becomes apparent when the green, modelled line in Figs. \ref{fig:lyman_BCD}, \ref{fig:lyman_A} best matches the data. While \texttt{VPFit} provides a statistical uncertainty for each fit, this is often unrealistic given the blending of most of the Lyman series transitions. We thus adopt a minimum uncertainty of 0.1 dex for those lines, although in the end only a limit is given.

Following this procedure, fits to $\ion{H}{I}$ 937, 930, 923, 920, 919, and 918 result in a total column density of $\log N_{\ion{H}{I}}^{\mathrm{A}} = 16.45$, which is consistent with the upper bound placed by the Lyman limit. Given the probability that all of the lines used in the fit are also to some degree contaminated by unrelated absorption, the values from individual components (given in Table \ref{table:all}) as well as the total are strictly upper limits. In reality, the $\ion{H}{I}$ is distributed in more sub-components than can be modelled (as the extra ones do not have corresponding metal absorption), which makes the fits appear inconsistent with the data. However, these extra components must not add much extra column, or else they would contradict the strict upper limit placed by the Lyman limit.

In contrast to system A, there are no unsaturated Lyman transitions corresponding to system B, causing difficulty in placing tight constraints on $N_{\ion{H}{I}}$. Due to this saturation, it is not possible to determine whether the $\ion{H}{I}$ appears to follow better the high-ions or low-ions. Nevertheless, we place upper limits on both BE ($\ion{H}{I}$ 917, 916) and BF ($\ion{H}{I}$ 920, 917), at $\log N_{\ion{H}{I}}^{\mathrm{BE,BF}}< 16.0$, with the $b$ parameter not well constrained (and largely unimportant). These values are over 2 dex lower than the main component and thus have negligible impact on the result.
By far the strongest component is BD/B$\beta,\gamma$. The saturation of the Ly-20 and Ly-21 lines as well as a total lack of flux below the corresponding Lyman limit (accounting for the flux error of 0.1 in this part of the spectrum) place a lower limit of $\log N_{\ion{H}{I}}^{\mathrm{BD}}> 17.9$. Assuming the $b$-parameter from the high-ions of 7\,\kms, the lack of damping in the red wing of Ly-$\alpha$ as well as the flux peak between systems B and C place a constraint of  $\log N_{\ion{H}{I}}^{\mathrm{BD}} \leq 18.35$, traced by the solid green line in Fig. \ref{fig:lyman_BCD}. The shaded region corresponds to column densities up to 0.3 dex away from this value, and it is clear that a value of  $\log N_{\ion{H}{I}}^{\mathrm{BD}} >18.35$ does not fit the data well. Varying the $b$- parameter as high as 20\, \kms~does not change this limit. The blue wing of component BD in $\ion{H}{I}$ 918, 917, 916, and 915 fits the observed flux best with a $b$-parameter of 20 \, \kms, which is significantly larger than that of both the corresponding low- and high-ions. While it is likely that in reality BD is formed of a number of smaller components with smaller $b$-parameters, and again allowing for contribution to the absorption from unrelated systems, it is with some confidence that we report the upper limit of $\log N_{\ion{H}{I}}^{\mathrm{BD}} \leq 18.35$. The components BA, BB and BC are unresolved and either contaminated or saturated throughout most of the Lyman series. Assuming the same $b$-parameters as the high-ions and using the constraining red wing of $\ion{H}{I}$ 918, we estimate a conservative combined limit of $\log N_{\ion{H}{I}}^{\mathrm{BA + BB + BC}}\leq 16.5$. Component BD greatly dominates the $N_{\ion{H}{I}}$ and, given the lower limit from the Lyman limit as well as the upper limits from the rest of the series, we estimate the total as $\log N_{\ion{H}{I}}^{\mathrm{B}}= 18.1 \pm 0.25$, where the probability distribution function is more likely flat than Gaussian, such that the true value is equally likely to lie anywhere within the uncertainty \citep{Prochaska2015}.

Systems C and D can be much more easily constrained by their unsaturated high-order Lyman transitions. For the central component of C we find the definite $\log N_{\ion{H}{I}}^{\mathrm{CB}}=15.53 \pm 0.05$, fixed predominantly by $\ion{H}{I}$ 930 and 926 (Ly-6, 7). We note that the best fit results when $b = 8$ \kms, which is lower than typically seen in $\ion{H}{I}$. However, this value is a true convergence of the program, rather than a fit by eye, and as such we assume it to be true. It also matches that of the metal lines. The high-ion component CB is mirrored by the low-ion C$\alpha$, but for consistency with systems A and B we use the capital letter notation for the $\ion{H}{I}$ in system C and D. The blended peripheral components are constrained to limits $\log N_{\ion{H}{I}}^{\mathrm{CA,CC}} < 14.0$ in the same transitions as well as Ly-$\alpha$, -$\beta$ and -$\gamma$. For system D we again apply the restriction of fits to $\ion{H}{I}$ 937, 926, 923, and 917 (Ly-$\epsilon$, 7, 8, and 12) which combine to give $\log N_{\ion{H}{I}}^{\mathrm{DB + DC}}= 15.9 \pm 0.1$. The red wing of Ly-$\alpha$ and the blue wing of $\ion{H}{I}$ 972, 937, 923, and 918 return values of $\log N_{\ion{H}{I}}^{\mathrm{DA}}= 14.35 \pm 0.1$ and $\log N_{\ion{H}{I}}^{\mathrm{DD}}= 14.0 \pm 0.1$.
Final measurement are presented in Table \ref{table:all}, and summed column densities for each system are given in Table \ref{table:totals}. 
 
\subsubsection{Metal column densities\label{subsec:metals}}

Systems A, B and C show significant absorption from both low- and high-ionisation states of various metal elements, including C, Si, and O, while system D shows a possible detection of weak absorption from $\ion{C}{II}$ and $\ion{Si}{II}$.
These systems are further broken up into a subset of velocity components, some of which are resolved while others are blended with one another.
Systems B and C also exhibit significant absorption from excited states of $\ion{C}{II}$ and $\ion{Si}{II}$. As it has a much larger $\ion{H}{I}$ column density and the largest corresponding columns of excited gas, we assign component B$_{\gamma}$ as that of zero relative velocity, which we keep consistent throughout the paper. 
We assume the instrument resolution of FWHM = 8\kms, which we verify by measuring the width of the single telluric lines. To fit Voigt-profiles to the absorption lines, we use the same method as described in \citet{Wiseman2017a}. To determine the number of components and their exact redshifts, we use unsaturated lines in high S/N regions of the spectrum, unaffected by blending from unrelated absorption. Following the criteria imposed on HIRES data by \citet{Prochaska2015}, we define an unsaturated line as one whose minimum flux is greater that 0.1 times that of the continuum, so as to avoid the limited hidden saturation that may occur. We select these initial lines separately for the singly and higher ionised species as the properties of the absorbing systems are significantly different. For each ionisation level, we fit the above-defined lines with \texttt{VPFit} and measure the broadening parameter, $b$, and redshift, $z$. We let the programme add components until the $\chi^2$ parameter is minimised. In some cases, such as B$\beta,\gamma$, this results in strongly blended components separated by only a few \kms, but this does not have a significant effect on the results. Once determined from these ideal lines, $z$ and $b$ are then fixed, and Voigt-profiles are fit to all of the other lines in that component and ionisation level in order to measure the column density $N$. An exception to this is the $\ion{O}{VI}$ detected in system A, which displays a unique absorption profile (Fig. \ref{fig:hiion_A}).
We show the low-ionisation fits in Figs. \ref{fig:loion_A}, \ref{fig:loion_B}, and \ref{fig:loion_C}, and those for the high-ionisation states in Figs. \ref{fig:hiion_A}, \ref{fig:hiion_B}, and \ref{fig:hiion_C}. Those of system D are combined and shown in Fig. \ref{fig:lonhi_D}. 

\subsection{Intervening Systems\label{subsec:analyse_intervening}}

Along with the absorption from the complex host galaxy system, the spectrum is populated with a vast number of intervening systems forming the Ly-$\alpha$ forest between Ly-$\alpha$ at the host redshift and its associated Lyman limit, corresponding to a redshift range of $2.32<z_{\textrm{abs}}<3.34$. A number of these show corresponding metal absorption, mostly in $\ion{C}{IV}$ and $\ion{Si}{IV}$, which suggests pockets of IGM gas that have been enriched with metals. The absorbers at $z=2.32$, 2.53, 3.157 and 3.300 also show significant absorption from singly-ionized species.

\section{Results\label{sec:results}}

The column densities measured by Voigt-profile fitting of each individual component are presented in Table \ref{table:all} of the Appendix, and a summary for each system is given in Table \ref{table:totals}. The four main systems are strikingly different in their absorption characteristics, and in the following section we present their physical properties revealed by the analysis.

%We also compare the abundances of different ionisation and excitation states of the same element. For simplicity, we use the following shorthand, using carbon as an example: $\ion{C}{II}/\ion{C}{IV} =\log(N(\ion{C}{II})/N(\ion{C}{IV}))$
\subsection{System A \label{subsec:A}}

System A, with $ \log N_{\ion{H}{I}}^{\mathrm{A}}<16.9$, is classed as a partial-Lyman limit system (pLLS). Its Lyman absorption is shown in Fig. \ref{fig:lyman_A}, and low and high ionisation metals in Figs. \ref{fig:loion_A} and \ref{fig:hiion_A} respectively. All $\ion{C}{IV}$ and $\ion{Si}{IV}$ transitions are affected by saturation in some or all of the seven main velocity components, which combine to give lower limits of $\log N_{\ion{C}{IV}}^{\mathrm{A}}>15.2$ and $\log N_{\ion{Si}{IV}}^{\mathrm{A}}\geq14.3$. The singly-ionised lines on the other hand are much weaker at $\log N_{\ion{C}{II}}^{\mathrm{A}}=13.32\pm0.02$ and $\log N_{\ion{Si}{II}}^{\mathrm{A}}=12.26 \pm0.03$, and are detected only in component A$\alpha$. The ratios of singly-to-triply ionised absorption are presented in Table \ref{table:ratios}, and are both around -2 dex, which indicates a very high degree of ionisation in the system. This assertion is reflected in oxygen, with the values of $\log N_{\ion{O}{VI}}^{\mathrm{A}}=14.8 \pm 0.05$ and $\log N_{\ion{O}{I}}^{\mathrm{A}}<12.05$. For all other ions we report non-detections, including both $\ion{N}{II}$ and $\ion{N}{V}$ despite good coverage in their respective wavelength regions. 

The aforementioned ionic ratios are an average calculated over the whole of system A. In reality the highly-ionised gas appears as a series of clouds located at different relative velocities along the line of sight while singly-ionised absorption is only detected in one component. In this component (A$\alpha$/AB) the ratios are less extreme although the difference between singly- and triply-ionised systems is still over 1 dex. In the other components, there is evidence that the ionisation is even stronger. The physical interpretation is that A$\alpha$ is a denser, slightly less-ionised pocket along a sightline of highly ionised gas covering over 300 \kms~in relative velocity. In general though, the fact that $\ion{H}{I}$ traces well the high-ions while the low-ions are only found in one narrow component suggests that the system is highly ionised, and that $\ion{H}{I}$ is simply tracing a much more dominant $\ion{H}{II}$ component. This implies that the metals too are mostly in the more highly ionised state, leaving too few to be detected in the singly ionised state.

\subsection{System B \label{subsec:B}}

With $17.9 < \log N_{\ion{H}{I}}^{\mathrm{B}} < 18.35$, system B has the largest column density of neutral hydrogen in the complex by nearly 2 orders of magnitude, yet the absence of evident damping of Ly-$\alpha$ (Fig. \ref{fig:lyman_BCD}) means it is also classed as a LLS. The singly-ionised species show one strong and three weaker components ranging over 100 \kms~(Fig. \ref{fig:loion_B}). The strongest, B$\gamma$, is effectively coincident with the high-ionisation component BD and is saturated in  $\ion{C}{II}~\lambda$1334 and $\ion{Si}{II}~\lambda\lambda$1190, 1193. High-ionisation lines, on the other hand, are much weaker than system A, and are unsaturated (Fig. \ref{fig:hiion_B}), allowing for tight constraints to be placed on their column density, and a comparison to the low-ions to be made. The $\ion{O}{VI}$ doublet is located in the Ly-$\alpha$ forest, and strong blending prevents an unambiguous detection. The ionic ratios  $\log(N_{\ion{Si}{II}}/N_{\ion{Si}{IV}})_{\mathrm{B}} =0.24\pm0.03$ and $\log(N_{\ion{C}{II}}/N_{\ion{C}{IV}})_{\mathrm{B}} >0.76$ are around 2 dex larger than system A, which implies that the ionisation is not as strong in system B. Similarly to system A, the ionic ratios vary substantially across system B: the gas associated with low-ion components B$\beta,\gamma$, and high-ion BD has a higher ratio of singly- to triply-ionised gas, while B$\delta$/BE appears much more strongly ionised. BA and BF appear to be wings of highly-ionised gas not detected in the low ions. Given that the low- and high-ion components do not trace each other exactly, it is not possible to calculate ratios for all individual components. Instead, we note that even within relatively small velocity ranges, ionic ratios appear to change on the order of a few tenths of a decade, which indicates a highly inhomogeneous ISM. Unfortunately, further analysis of the ionisation state of the system is limited due to the contamination by unrelated absorption in areas of the spectrum containing useful species such as $\ion{Si}{III}$ and $\ion{N}{I}$. Along with the large uncertainty on $N_{\ion{H}{I}}$, this restricts our ability to constrain the metallicity of the system. With an unsaturated detection of $\ion{O}{I}$, however, we can at least place some constraints that are independent of ionisation. This is because the neutral states of O and H are coupled by charge-exchange reactions \citep{Prochaska2015}, so they are likely to have similar ionisation fractions such that the relative abundance of the neutral state, $\{\ion{O}{I}/\ion{H}{I}\}_{\mathrm{B}}=\log (N_{\ion{O}{I}}/N_{\ion{H}{I}})_{\mathrm{B}}-\log (N_{\mathrm{O}}/N_{\mathrm{H}})_{\sun}$, represents the overall relative abundance,  [O/H]$_{\mathrm{B}}=\log (N_{\mathrm{O}}/N_{\mathrm{H}})_{\mathrm{B}}-\log (N_{\mathrm{O}}/N_{\mathrm{H}})_{\sun}$. Following this methodology results in -1.3 < [O/H]$_{\mathrm{B}}$< -0.8. We note, however, that this assumption may break down at the low $N_{\ion{H}{I}}$ of this absorber, as well as at the high X-ray fluxes observed in GRB 080810 \citep{Page2009}. 
\begin{table}
\centering
\small
\caption{Total column densities in the four main systems of the host complex.}
\label{table:totals}
\begin{tabular}{c c c c c}
\hline\hline\noalign{\smallskip}
 & A & B & C & D \\
\hline\noalign{\smallskip}
  Ion& \multicolumn{4}{c}{$\log N$}\\
\hline\noalign{\smallskip}
$\ion{H}{I}$&$\leq$16.9&$18.1\pm0.25$\tablefootmark{a}&$15.53\pm0.05$&$15.97\pm0.1$\\
$\ion{C}{II}$& $13.32\pm0.02$&$>14.5$ & $13.50\pm0.02$&$12.66\pm0.05$\\
$\ion{C}{II}$*& $<12.35$&$>14.5$ & $13.11\pm0.02$& $<12.55$\\
$\ion{C}{IV}$&$>15.2$ & 14.03$ \pm 0.01$& $13.70\pm0.07$& <12.05\\
$\ion{N}{II}$& $<13.35$& $14.11\pm0.02$& $<12.75$& $<12.15$\\
$\ion{N}{V}$&$<12.0$ & $<12.3$& $<13.05$& $<11.85$\\
$\ion{O}{I}$& $<12.05$& $13.73\pm0.03$&$ <11.2$& $<12.5$\\
$\ion{O}{VI}$& $14.81\pm0.05$& -&- &- \\
$\ion{Al}{II}$& $<11.3$&$>13.0$ & $11.66\pm0.04$&$\leq11.2$ \\
$\ion{Si}{II}$&$12.26\pm0.03$ & $13.87\pm0.02$& $13.04\pm0.03$& $\leq 11.65$\\
$\ion{Si}{II}$*& $<12.25$& $12.94\pm0.03$& $11.7\pm0.1$& $<12.4$\\
$\ion{Si}{IV}$& $\geq14.3$& $13.68\pm0.01$& 12.86$\pm0.03$& $<11.3$\\
$\ion{Fe}{II}$& $<12.65$& $<12.70$&<12.75 &$<12.5$ \\
\hline
\end{tabular}
\tablefoot{(a) PDF is not Gaussian, but more likely flat.}
\end{table}

System B shows strong absorption from the unstable fine structure transitions $\ion{C}{II}$* and $\ion{Si}{II}$*. Both the $\ion{C}{II}$ and $\ion{C}{II}$* absorption is saturated, preventing us from calculating the ratio of excited to ground state $\ion{C}{II}$. This saturation is very mild, and at such high resolution and S/N it does not significantly affect the values obtained from the Voigt-profile fit. We conclude the excited fraction is around 50\%. $\ion{Si}{II}$ and $\ion{Si}{II}$*, on the other hand, are not saturated in some transitions, allowing an exact ratio to be calculated at $\log(N_{\ion{Si}{II}\mathrm{*}}/N_{\ion{Si}{II}})_{\mathrm{B}}=-0.93 \pm 0.04$, or $\sim$ 10\%. Interestingly, $\ion{O}{I}$* $\lambda$ 1304 and $\ion{O}{I}$** $\lambda$ 1306 are not detected unlike e.g. GRB 050730 \citep{Prochaska2006b} where $\ion{C}{II}$* and $\ion{Si}{II}$* are also seen.

\subsection{System C \label{subsec:C}}

System C has the lowest $\ion{H}{I}$ column density of the host complex with $\log N_{\ion{H}{I}}^{\mathrm{C}}=15.53 \pm 0.05$ (Fig. \ref{fig:lyman_BCD}), yet shows strong metal absorption (Fig. \ref{fig:loion_C}), even from the low-ionisation lines of C, Si and Al, suggesting that it has also been significantly enriched with metals. We compare the relative abundances of $\ion{Si}{II}$ and $\ion{H}{I}$: $\{\ion{Si}{II}/\ion{H}{I}\}_{\mathrm{C}}=2.00$ is over 1.5 dex more than that of system B, which is constrained to $0.06 < \{\ion{Si}{II}/\ion{H}{I}\}_{\mathrm{B}}<0.56$. This could be explained if the metal content was the same but system C was much more ionised than system B, such that $N_{\ion{H}{I}} / N_{\mathrm{H}} \ll 1$, a hypothesis supported by the lack of an $\ion{O}{I}$ detection in C. However, the high-ionisation lines in system C are weaker than system B (Fig. \ref{fig:hiion_C}), such that the ionisation strength appears not to be particularly high.
We make qualitative inferences on the nature of systems B and C in Sect. \ref{subsec:discuss_bc}.
%However, we take a close look at the system at $z=3.55$ towards PKS$2000-330$, analysed in \citet{Prochter2010}, comprising three LLS very similar to those in the host complex of \object{GRB 080810}. Of these, our system B most resembles their subsystem C, and our system C is similar to their subsystem B. Indeed, our ionisation ratios for silicon, carbon and oxygen agree to within $\sim$0.3 dex. Fig. 6 of that paper shows ionic ratio curves as a function of the ionisation parameter $U$, from which it can be seen that the uncertainty on $U$ is most likely larger than the difference between the ratios from the two studies. Their results of $U=-2.4$ and $-2.8$ for their subsystems C and B are therefore reasonably likely to represent those in our systems B and C respectively. The very fact that $U$ is comparable for systems B and C is remarkable, given the difference in $N_{\ion{H}{I}}$. An explanation for the ionisation characteristics is presented in Section \ref{sec:discussion}.

Fine structure lines of $\ion{C}{II}$* and $\ion{Si}{II}$* are also detected in system C, albeit significantly weaker than in B. All but two $\ion{Si}{II}$* transitions are contaminated by unrelated absorption: $\ion{Si}{II}$* $\lambda1264$ is detected, but with a small column density, such that the non-detection of $\ion{Si}{II}$* $\lambda1309$ is consistent. Ratios of $\log(N_{\ion{Si}{II}\mathrm{*}}/N_{\ion{Si}{II}})_{\mathrm{C}}=-1.34$ and $\log(N_{\ion{C}{II}\mathrm{*}}/N_{\ion{C}{II}})_{\mathrm{C}}=-0.39$ indicate less excitation than in B. 

\begin{table}
\centering
\small
\caption{Ionic and excitation ratios for the four components in silicon and carbon, given in the form $\ion{X}{A}/\ion{X}{B} = \log(N_{\ion{X}{A}}/N_{\ion{X}{B}})$. These ratios are derived for entire systems; the values for individual components vary somewhat. }
\label{table:ratios}
\begin{tabular}{l c c c c}
\hline\hline\noalign{\smallskip}
 &  A&B&C&D\\
\hline\noalign{\smallskip}
 Ratio & \multicolumn{4}{c}{$\log(N_{\ion{X}{A}}/N_{\ion{X}{B}})$}\\
\hline\noalign{\smallskip}
$\ion{C}{II}/\ion{C}{IV}$&$<-1.9$& $\geq$0.8&-0.05$\pm$ 0.1&$>0.5$\\ 
$\ion{Si}{II}/\ion{Si}{IV}$&$\leq -2$&$0.24 \pm 0.03$&0.22$\pm0.04$&-\\
$\ion{C}{II}*/\ion{C}{II}$&$<-0.97$& $\approx 0$&$ -0.39\pm0.03$& <0.11\\
$\ion{Si}{II}*/\ion{Si}{II}$&$<0$&$-0.93\pm0.04$&$-1.34\pm0.1$&-\\
$\ion{O}{I}/\ion{O}{VI}$&$<-2.75$&- &- &- \\

\hline
\end{tabular}

\end{table}

\subsection{System D \label{subsec:D}}

System D has an $N_{\ion{H}{I}}$ that is 0.4 dex larger than system C, yet there is an almost total dearth of metal absorption apart from a marginal ($\sim$3 $\sigma$) detection of $\ion{C}{II}$. Standout measurements include the lack of any highly-ionised metal lines, with upper limits of $\log N_{\ion{C}{IV}}^{\mathrm{D}}<12.05$ and $\log N_{\ion{Si}{IV}}^{\mathrm{D}}<11.05$. This places a limit of $\log (N_{\ion{C}{II}}/N_{\ion{C}{IV}})_{\mathrm{D}} >0.5$ dex, which points towards a similar or lower ionisation fraction compared to system C. The relative abundance, on the other hand, is much lower: $\{\ion{C}{II}/\ion{H}{I}\}_{\mathrm{D}} = 0.33$ compared to  $\{\ion{C}{II}/\ion{H}{I}\}_{\mathrm{C}} = 1.69$.  $\ion{Si}{II}$ is very weak, but may be evident in the transition at $1260 ~\AA$. This part of the spectrum, at $5478 ~\AA$ in the observer frame, is contaminated by $\ion{C}{IV}$ $\lambda$~1550 absorption at $z=2.534$, and no absorption from the other available Si lines is detected above the level of the noise. If we assume the null hypothesis that there is no $\ion{Si}{II}$ absorption, the upper limit we measure at the position of the $1190~\AA$, $1304~\AA$ and $1526~\AA$ transitions is consistent with the column density measured from the tentative $1260~\AA$ feature, $\log N_{\ion{Si}{II}}^D \leq 11.65$. We thus estimate a conservative upper limit of  $\{\ion{Si}{II}/\ion{H}{I}\}_{\mathrm{D}} \leq 0.24$, consistent with that of $\ion{C}{II}$. The wavelength region around $\ion{Si}{III}~\lambda$ 1206 ($5424~\AA$ in the observed frame) is contaminated by a Ly-$\alpha$ line at $z=3.313$. Based on the loose assumption of a similar ionisation level to that in C, we infer a difference in metallicity of around 1.5 dex, which is vast when taking into account the small (160 \kms) difference in relative velocity.

\subsection{Imaging and emission spectrum\label{subsec:imaging}}

\begin{figure*}
    \centering
\includegraphics[width=18.5cm]{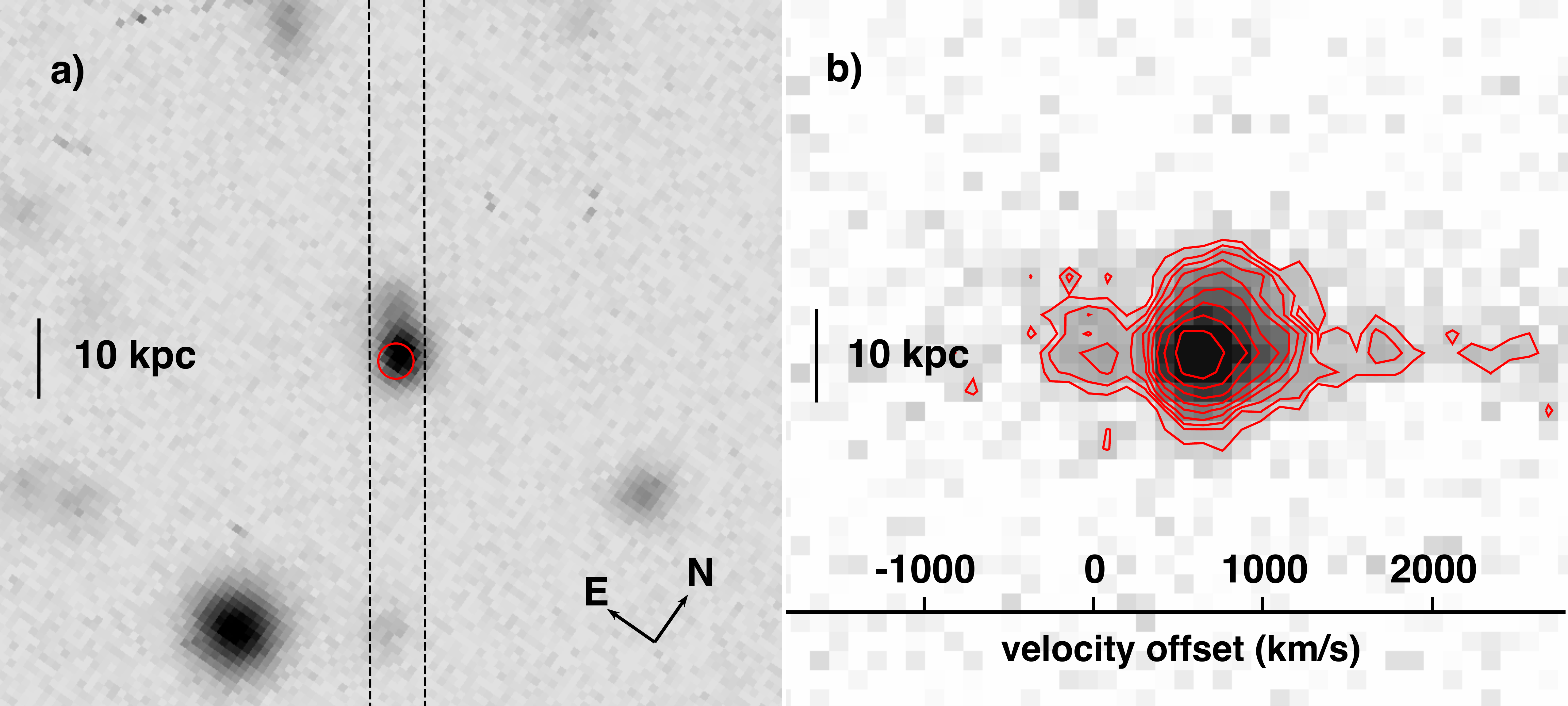}%{im2dspec2blc2.eps}
\caption{\label{fig:img} \small a) VLT/FORS2 $R$-Band image with the GRB location marked with the red circle. The slit position used for the spectral observations is marked with vertical dotted lines; b) Keck/LRIS low resolution 2D spectrum around the wavelength of Ly-$\alpha$ at $z\approx3.35$. The contours correspond to the relative flux of Ly-$\alpha$ emission. The velocity scale is centered at the location of absorption component B$\gamma$. }
\end{figure*}

A galaxy is detected at the GRB afterglow location in all of the 
images we took of the GRB field. The deep VLT image is shown in the left panel of Fig. \ref{fig:img}; 
the host exhibits an extended morphology with a bright compact region accompanied by a 
a fainter secondary component offset 0.8$\arcsec$ to the northeast.  A 
strong drop in the flux in the $g$- and $B$-bands suggests that this 
source is indeed the host galaxy system.  Our spectroscopy confirms 
this, revealing a strong detection of redshifted Ly-$\alpha$ centered 
at 5305 \AA ~($z=3.36$).

The host is marginally resolved in both position and velocity space.  An 
extraction of the spectrum at the position of the primary component seen 
in the imaging shows two marginally-resolved peaks in wavelength space 
separated by $\sim$13 \AA~(observer-frame): 
one at our defined central redshift of $z=3.35$, and a second dominant peak redshifted by $\sim$700 \kms.  There is no 
spatial offset evident between these components in the 2D spectrum.  Our 
LRIS longslit was oriented to include both the ``primary'' component of 
the host system and the secondary source to the NE, so this secondary 
source is also visible on the trace although it is partially blended 
with emission from the ``primary''. It shows bright Ly-$\alpha$ 
emission redshifted by approximately +250 \kms~relative to the strongest 
emitter, placing it at +1000 \kms~ redder than our systemic redshift.

Given the equivalent redshifts, it is natural to associate the strongest 
emitter in the host spectrum with component A in the afterglow 
absorption spectrum; and associate the blueshifted 
(spatially-coincident) emission with component B (Fig. \ref{fig:img}b).  The spatially-offset 
source and its red tail have no counterpart in absorption, while 
sources C and D have no distinct counterparts in emission (however, the 
low resolution of the host spectrum means that any such counterpart 
would likely be hopelessly blended with emission from B). If the GRB occurred in component B, then component A must be 
in the foreground and therefore must be infalling towards the host given 
its higher redshift---most likely due to an active merger (the lack of a 
detectable spatial offset puts them within a distance of $\sim$5 kpc in 
projection, so unless they are an improbable line-of-sight projection 
the merger must be actively ongoing.)

\begin{table}
\centering
\small
\caption{Optical and NIR photometry of the host system.}
\label{table:photometry}
\begin{tabular}{c c l}
\hline\hline\noalign{\smallskip}
Band & Magnitude \tablefootmark{a} & Telescope/Instrument \\
\hline\noalign{\smallskip}
 $B $    &$ 24.36\pm 0.09$&Keck-I/LRISb \\
   $g$   &$    24.13\pm 0.07$&Keck-I/LRISb\\
  $ r $   &$   23.33 \pm0.10 $&GTC/OSIRIS\\
   $R$   &$    23.25\pm 0.04$&VLT/FORS2\\
  $ R $  &$    23.42\pm 0.06$&Keck-I/LRISr\\
  $ i  $  &$   23.22 \pm0.06$&GTC/OSIRIS\\
  $ z$   &$    23.46\pm 0.15$&Keck-I/LRISr\\
 $  J$   &$    23.44\pm 0.22$&Keck-I/MOSFIRE\\
  $ Ks $&$     22.23\pm 0.12$&Keck-I/MOSFIRE\\
   3.6&$     23.57\pm 0.07$&Spitzer/IRAC\\
\hline
\end{tabular}
\tablefoot{(a) AB system, not corrected for foreground reddening.}
\end{table}
We fitted all of our existing broadband photometry (presented in Table \ref{table:photometry}) with the Galaxy Builder photometric SED
analysis software (first described in \citealt{Perley2012,Perley2013}).  All
of the data points except for those in the K-band and \emph{Spitzer} 3.6 $\mu$m are located at redshifts blueward of the redshifted
Balmer break and the $K_s$-band is strongly contaminated by nebular emission
from [OIII] and H$\beta$, so it is difficult to constrain the presence of
older stars in the population.  However, it is clear that the
star-formation rate must be very large ($\sim10^2 M_{\odot}$ yr$^{-1}$) in
order to explain the luminous UV emission and $K_s$-band nebular excess.
 Extinction is modest ($A_V \sim 0.4$ mag) and the inferred stellar mass (approximately $\sim$3$\times$10$^9 M_{\odot}$) is moderate, similar to the median for GRB hosts generally \citep{Perley2016b}.

 \section{Characterising the absorbing systems\label{sec:geometry}}
Using the chemical, ionic and kinematic properties presented in the previous section, we now classify the four distinct systems at the highest redshift along the extremely rich line of sight to \object{GRB 080810} and build a picture of the entire complex, which we summarise in the cartoon Fig. \ref{fig:scenario}.

\subsection{Systems B \& C\label{subsec:discuss_bc}}
There are several factors that lead us to believe that system B is the one nearest to the explosion site of \object{GRB 080810}. The most compelling is that it is by far the largest reservoir of neutral hydrogen of the four systems studied, a characteristic trait of the LOS probed by GRBs through their host galaxies. GRBs are associated with massive stars, and while they typically fully ionise their immediate surroundings, one expects the LOS to probe other regions of cool, dense and predominantly neutral gas necessary for star formation inside the host galaxy. Although system B falls well below the average $N_{\ion{H}{I}}$ for a GRB (in \citealt{Cucchiara2015} the median is $\log N_{\ion{H}{I}}=21.5$), its value is nearly two orders of magnitude larger than systems A, C, and D combined. 

\begin{figure}
    \centering
\includegraphics[width=9.0cm]{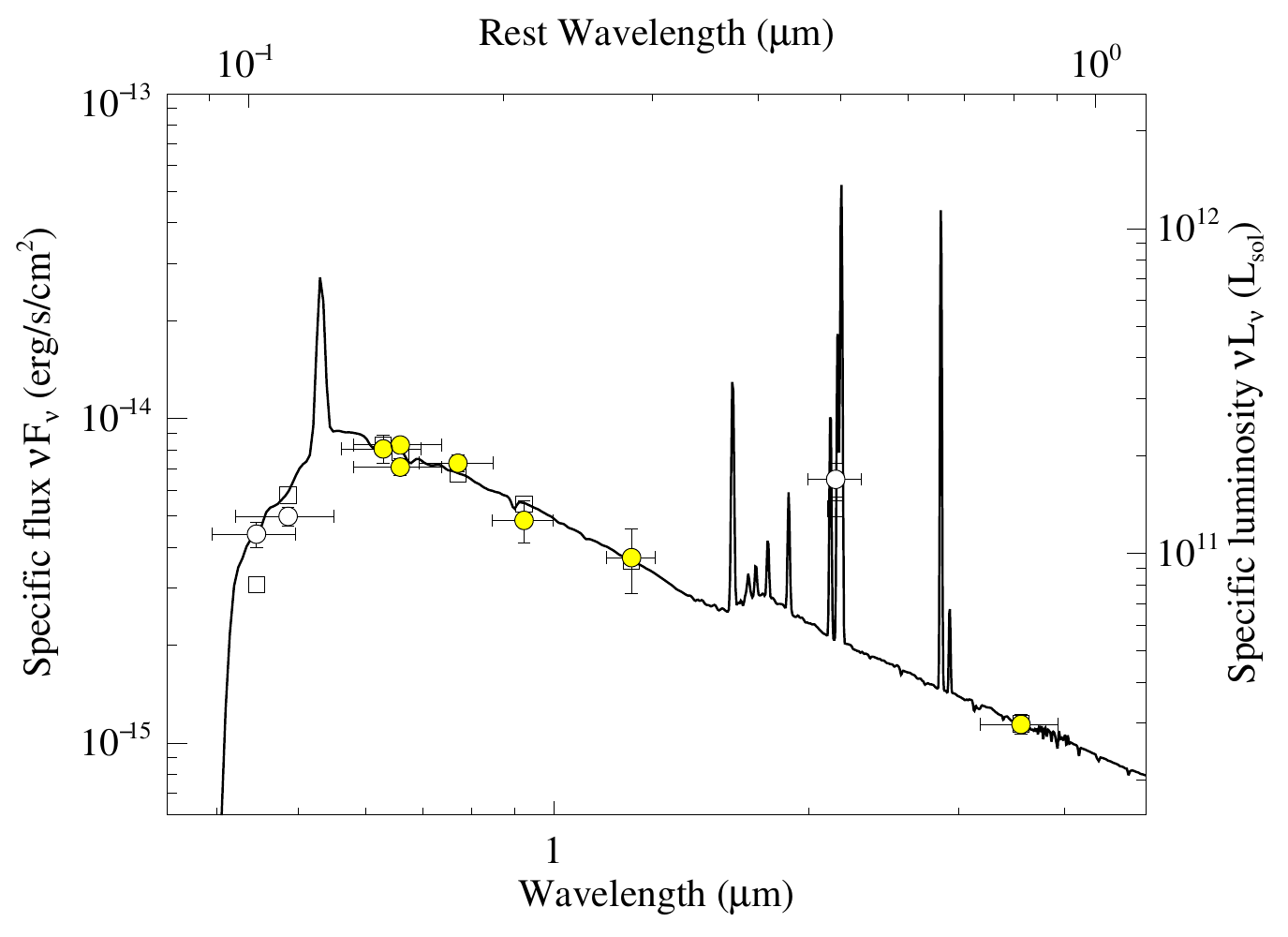}%{sedfit.jpg}%{im2dspec2blc2.eps}
\caption{\label{fig:sed} \small The spectral energy distribution of the host galaxy. We have obtained thorough multi-band photometry of this host as part of the SHOALS survey, showing it to be a remarkably luminous (MUV $\sim -$23 mag) Lyman-break galaxy. A notable excess in the $K_s$-band photometry above the SED model suggests very strong, high equivalent-width [OIII] emission. Clear circles show bands not used in the SED fit due to Ly-$\alpha$ forest or nebular contamination). Empty squares represent the value of the model integrated over the width each band, and show a relatively good fit. }
\end{figure}

Secondly, we refer to the strong absorption by the fine-structure transitions of $\ion{C}{II}$ and $\ion{Si}{II}$, often used as strong evidence that the physical location of the cloud is near to the GRB \citep{Prochaska2006b,Vreeswijk2013,DElia2014,Friis2015}, as the intense UV radiation from the burst excites these unstable transitions. System C however also exhibits strong fine-structure absorption, and as such could also be close to the GRB. This observation is similar to the results of \citet{Savaglio2012} on the pair of interacting galaxies forming the host of \object{GRB 090323}, which both exhibit absorption from excited lines. It is known that lines may also be excited by UV radiation from a strong starburst \citep{Pettini2002,Christensen2011}, and the interpretation presented for \object{GRB 090323} is that substantially increased star-formation caused by the interaction caused the excitation. Systems B and C may similarly be affected by intense background UV flux from the highly star-forming environment, but they would have to both be very close to the young, hot stars. It is more likely that the majority of the excitation is caused by the GRB in both cases, as this can occur at distances up to a few kiloparsecs from the burst. We note that the relative fraction of $\ion{Si}{II}$ in the excited state is 0.3 dex higher in system B, and a similar pattern is evident in $\ion{C}{II}$. Assuming that both systems are exposed to the same background radiation field, we suggest that the extra excitation seen in B could be down to a higher influence of UV pumping from the GRB itself, and thus that it is located closer to the site of the burst. 

System C is 170 \kms~blueshifted from the central velocity component of B. However, it shows a fairly similar chemical composition. The similarities in the ionisation and excitation characteristics of B and C suggest that they are irradiated by a similar source, namely the GRB plus some UV background from nearby OB stars. Typically it is expected that of two clouds exposed to similar ionising fields, the one with lower $N_{\ion{H}{I}}$ will be more strongly ionised, as self-shielding is less efficient. For its very low $\log N_{\ion{H}{I}}=15.53$, the ionisation in C appears remarkably weak, based upon $\ion{C}{IV}$ and $\ion{Si}{IV}$ measurements (Table \ref{table:ratios}) which suggest it could be similar to that in B. With comparable ionisation but much less $\ion{H}{I}$, we infer that the total hydrogen content of system C is much lower than B. Thus, given that the metal detections are still strong, we assume metallicity is likely similar or even higher. Along with the negative relative velocity of nearly 200 \kms , we are led to propose that system C may trace an outflow driven by the intense star formation and associated supernovae in the galaxy in which the GRB exploded. 
Although the ionisation is relatively weak compared to system B, the system is still strongly ionised, with the $\ion{H}{I}$ tracing the high-ion components CA and CC that are not seen in low-ions. This observation is consistent with other observations of strong ionisation in galactic outflows \citep{Pettini2002,Grimes2009,Heckman2015,Chisholm2016}.%The high metallicity implies a high efficiency of metal expulsion during a previous, or the ongoing,  epoch of intense star formation, with the metallicity of the remaining gas (i.e. that traced by B) kept lower by the inflow of metal-poor gas. On the other hand, the outflow could be more metal rich than the ambient ISM as it contains some of the insides of the exploded stars driving the wind, and if it has a relatively small mass-loading factor.

\subsubsection{The lack of an Fe detection\label{subsubsec:depletion}}
In dense regions of the ISM, metals are often found to condense out of the gas phase and be locked into dust grains, an effect known as dust depletion (e.g. \citealt{Savage1996,Vladilo1998,Savaglio2003,DeCia2013}). Furthermore, certain refractory elements such as Fe, Ni and Ti are known to be much more heavily depleted than volatile elements such as Zn, S and P. The relative depletion strengths of many elements have been well characterised by \citet{Jenkins2009} and most recently \citet{DeCia2016}, such that with a combination of measurements for volatile and refractory elements it is possible to constrain the dust content of an absorbing system, which has been done for a sample of GRB host galaxies by \citet{Wiseman2017a}. In that work, column density measurements for a minimum of four elements were required to constrain the depletion, but it is in practice possible to gain an insight into the dust content from just one volatile and one refractory element (e.g. \citealt{Vladilo2011,DeCia2013}). In the case of system B of \object{GRB 080810}, we do not detect any strongly refractory elements. Furthermore, due to ionisation the measurements of certain states cannot be treated as fully representative of the elements as a whole, meaning it is not possible to measure the dust depletion directly. However, iron and silicon have very similar ionisation potentials as well as solar abundances, so given the strong detection of $\ion{Si}{II}$ ($N_{\ion{Si}{II}}^{\mathrm{B}} = 13.94$) we would expect a similar amount of $\ion{Fe}{II}$. Instead, we measure only an upper limit of $\log N_{\ion{Fe}{II}}^{\mathrm{B}}<12.75$. Assuming that iron follows a similar ionisation pattern to silicon, we can place a tentative constraint of [Si/Fe]$_{\mathrm{B}}$>0.9. If this value is due exclusively to dust, it means that system B shows stronger dust depletion than any of the 19 GRB-DLAs analysed in \citet{Wiseman2017a}, at a level similar to that of the average Milky Way sight-line \citep{Jenkins2009,DeCia2016}. There are many caveats to this analysis. It requires that the nucleosynthesis history of this galaxy, at $z=3.35$, is similar to that in the solar neighbourhood, as well as that our assumptions about ionisation fractions are correct. We note that three separate studies have detected little or no dust in the afterglow SED: $A_V=0$ by \citet{Page2009}; $A_V = 0.16\pm0.02$ by \citet{Kann2010}, and $A_V<0.35$ by \citet{Schady2012}. We also find the SED of the host gives an $A_V$ of only $\leq 0.5$ mag. Such a small amount of extinction is not expected for such a large amount of depletion, although these two measures of dust are indeed often inconsistent with each other \citep{Watson2006,Friis2015,Wiseman2017a}. Although there are many uncertainties in play, it is possible that some of this Fe/Si discrepancy could be due to dust. Such a detection could suggest that the gas has been ionised only recently, the cores of dust grains remaining intact. 

Another possible reason for the lack of Fe is that at this redshift we expect a strong enhancement of the abundance of the alpha elements (usually denoted [$\alpha$/Fe]), such that the relative abundances of of Si, O, C compared to Fe are positive. However, typically these enhancements are of a maximum of $\sim$ 0.3 dex in high redshift absorption systems \citep{Dessauges-Zavadsky2006, Rafelski2012,DeCia2016}. Furthermore, we take a look at the relative abundance of another non-alpha element with similar ionisation potential to Fe: Aluminium. The potentials for the first two ionisation states of Al are similar to those for Fe and Si, so the value $\{\ion{Al}{II}/\ion{Si}{II}\} \approx 0.06$ suggests there is very little $\alpha$ enhancement.

The depletion characteristics of Al are poorly understood, but is usually described as a refractory element \citep{Phillips1982,Prochaska2002a}. The similarity with the Si abundance thus suggests that the depletion is  likely small, in contrast to the inference from the Si/Fe ratio. This could be explained by an abnormally large Al/Fe ratio, as seen in other GRBs \citep{DElia2014,Hartoog2015}, for which proton capture processes were highlighted as a possible cause. Were this the case in GRB 080810, then this would prevent the Al/Si ratio from ruling out $\alpha$ enhancement being prominent.

While we cannot discount the possibility that all of these deviations from expected abundance patterns are primarily caused by the effects of strong ionisation, it remains that the non-detection of iron is particularly puzzling.

\subsection{System A\label{subsec:discuss_a}}
The absorption system A has a mean velocity of $+690$ \kms~relative to system B, consistent with the relative velocity seen in Ly-$\alpha$ emission. The implication is that system A lies in the foreground and is falling towards, and likely merging with, the host (including systems B and C). System A shows drastically different chemical and kinematic properties to system B. It is almost totally ionised, and shows absorption from higher ionisation states over a 320 \kms~range in relative velocity, which is traced by relatively weak neutral hydrogen. This velocity range is consistent with that seen at intermediate impact parameters ($\sim$ 30 kpc) in the CGM of a sample of Lyman break galaxies (LBGs) at a similar redshift by \citet{Steidel2010}. The presence of such highly ionised gas, including strong $\ion{O}{VI}$ absorption, points to system A being a sight line through the CGM of the foreground galaxy. Indeed, the high ionisation and wide velocity structure is similar to that observed in the CGM of star-forming galaxies in the COS-Halos sample at $z\approx0.2$ \citep{Werk2016,Prochaska2017}. However, the imaging data presented in Sect. \ref{subsec:imaging} imply that the LOS from the GRB passes within $\lesssim$5 kpc of the centre of this foreground galaxy. This is over an order of magnitude smaller than the impact parameters for the \citet{Steidel2010} sample, as well as the COS-Halos sample and the $z\lesssim1$ LLSs by \citet{Lehner2013}. The fact that the LOS passes so close to the galaxy centre yet incurs such little neutral gas is unprecedented, and implies a very compact system --- whereas in the aforementioned samples one would expect at least a sub-DLA for such a sight line, here we see the typical signature of the CGM. 

\subsection{System D \label{subsec:discuss_d}}
There is only a $\sim$150 \kms~difference in relative velocity between systems C and D, yet they have strikingly different chemical properties. The near-total lack of metal absorption in D contrasts starkly with the comparatively metal-rich system C, and the lack of $\ion{C}{IV}$ and $\ion{Si}{IV}$ suggest a small ionised fraction. Together, these characteristics point to D being a cooler, metal poor cloud that is part of an inflow from the IGM. Although the marginal detection of carbon means that this is likely not primordial gas, it is much less enriched than that in the other systems, which have been processed in stars. The metals present could originate from mixing with more metal rich gas ejected during previous epochs of intense star formation from nearby galaxies (see  \citealt{Fraternali2016} for a review).

\begin{figure}
   
   \centering
   \includegraphics[width=9.0cm]{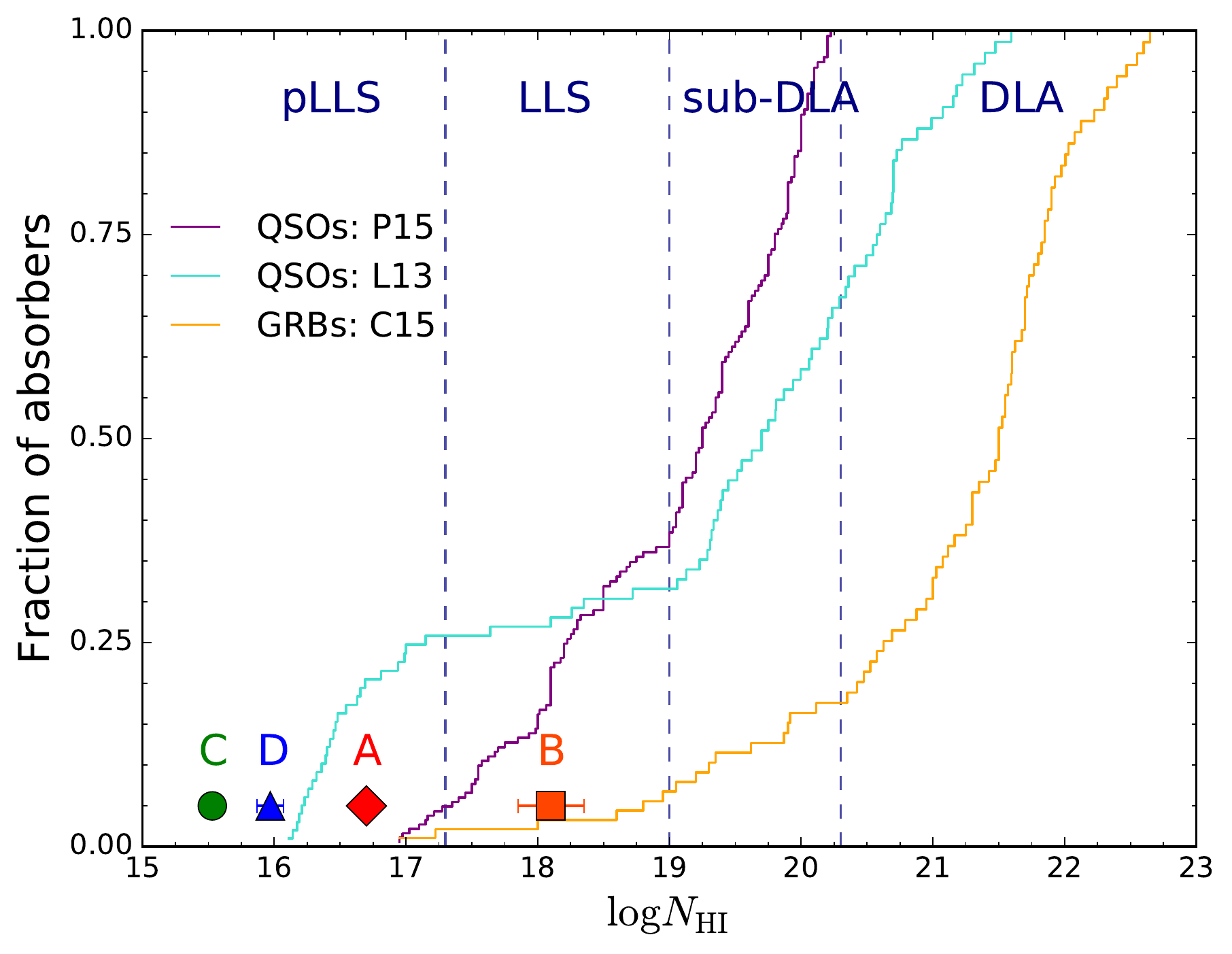}
   
   \caption[blah]{\label{fig:nh} \small The neutral hydrogen column in the four systems (coloured shapes) compared to cumulative distributions of $N_{\ion{H}{I}}$ from three samples of Ly-$\alpha$ absorbing systems: GRB host galaxies (\citealt{Cucchiara2015}; C15, to which we have added GRBs 120119A and 141028A from \citealt{Wiseman2017a}), and QSO absorbers (\citealt{Lehner2013}; L13, \citealt{Prochaska2015}; P15).}
   \end{figure}

\begin{figure*}
  \centering
   \includegraphics[width=18cm]{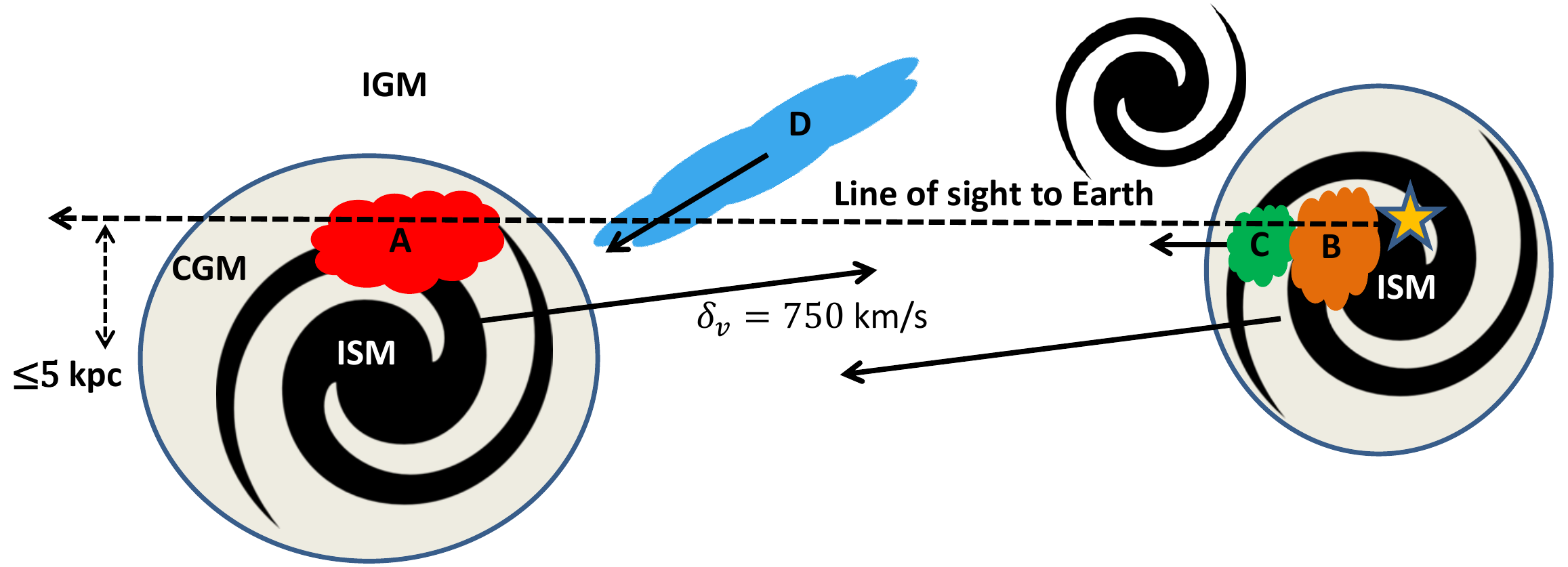}
   
   \caption{\label{fig:scenario} \small The scenario derived for the geometry of the host system. The GRB is represented by a yellow star, with the four absorption systems which trace strikingly different phases of gas coloured according to their apparent redshift in the absorption spectrum. The 5 kpc offset comes from the size of the foreground galaxy in the image along with the uncertainty on the GRB position. The extra component seen in the imaging data (Fig. \ref{fig:img}), here shown between the two galaxies, does not intersect the line of sight, and could be located at any horizontal position in this diagram.}
\end{figure*}
\section{Discussion \label{sec:discussion}}

\subsection{Conditions in the interacting galaxies \label{subsec:properties}}

The analysis of the four main components paints a picture of a complex, compact environment that plays host to \object{GRB 080810}. Typically, denser gas is expected to be found at smaller impact parameters ($R_{\perp}$), a trend seen in \citet{Werk2012}, \citet{Lehner2013} and \citet{Prochaska2017}. In the sample of LBGs presented in \citet{Steidel2010}, there is also a strong decrease of absorption line strengths of Ly-$\alpha$ and metals with increasing impact parameter. As shown in Fig. \ref{fig:nh}, systems A, C and D have some of the smallest column densities of neutral hydrogen ever observed at the host redshift of a GRB (compared to objects in the sample of GRB-DLAs by \citealt{Cucchiara2015}), with B also at the very low end of the distribution. We also contrast these column densities with two samples of QSO absorption line systems that probe the CGM, namely \citet{Prochaska2015} and \citet{Lehner2013}. Whereas the QSO absorber samples typically probe gas at impact parameters of $R_{\perp}\sim 10-100$ kpc, the sight line to \object{GRB 080810} passes within 5 kpc, yet it is evident that the $N_{\ion{H}{I}}$ is at the lower end of these populations as well. To find such a low $N_{\ion{H}{I}}$ in a system that is both a GRB host and lies at a small radius is highly unusual. 

Systems A and B display comparable total metal column densities, yet the fractions of C and Si in higher ionisation states are over two orders of magnitude larger in A than in B. This is reflected in the observed neutral hydrogen column density, which is 1.4 dex smaller in A. This result is consistent with that from \citet{Prochaska2015}, who found that the ionised fraction is higher in LLSs with lower $N_{\ion{H}{I}}$ (their Figure 9). We note that the sample presented in that paper has a lower $\log N_{\ion{H}{I}}$ cut-off of 17.3, and that trends evident there may not apply at the much lower column densities probed by A, C and D.
The vast difference in ionisation fractions and large velocity widths of systems A, B, C, and D imply highly turbulent media, reminiscent of that found in CGM at larger ($\sim$100 kpc) radii (e.g. \citealt{Crighton2013}). We emphasise that in star-forming galaxies at high redshift there is a dynamic interplay between the the ISM and CGM (see e.g. \citealt{Steidel2010}), and the boundaries are largely undefined. What is clear is that these are compact galaxies, with relatively small amounts of gas being probed. The gas observed is much more diffuse than expected for lines of sight passing so close to centres of star-forming galaxies.

While low $\ion{H}{I}$ column densities are rare in GRBs, a common trait exists among the small number of such objects, namely that multiple absorbing components separated by large (500 - 5000 \kms) velocities are often detected, as presented for GRB 080810 in this work. \object{GRB 021004} ($\log N_{\ion{H}{I}}=19.0$), \object{GRB 060607A} ($\log N_{\ion{H}{I}}=16.8$), and \object{GRB 080310} ($\log N_{\ion{H}{I}}=18.8$) were all analysed in \citet{Fox2008}, and \object{GRB 060605} ($\log N_{\ion{H}{I}}=18.9$) by \citet{Ferrero2009}. Those papers discuss the nature of the observed high-velocity components: whether they are winds originating from the GRB progenitor, are outflows on galactic scales, or are unrelated intervening systems. However, our analysis of the imaging and late-time spectroscopy of the host of \object{GRB 080810} strongly suggests that the two largest absorbing systems correspond to two very compact, interacting galaxies.

\subsection{GRBs and SNe in interacting systems\label{subsec:mergers}}

The link between mergers and intense star-formation has been studied for decades (e.g. \citealt{Sanders1988,Bell2006,Robaina2009}; see \citealt{Bournaud2011} for a review), and such events are a key ingredient in the theory of galaxy evolution \citep{White1978}. Given that core collapse SNe and GRBs are associated with massive stars, it is reasonable to expect a link between systems where mergers have caused starburst activity and thus a young and massive stellar population, and GRBs \citep{Lopez-Sanchez2009,Chen2011}, and we propose that \object{GRB 080810} could be one such system. There are other examples of potentially merging systems hosting cosmic explosions, detected in absorption line profiles such as \object{GRB 090323} \citep{Savaglio2012} and \object{GRB 100219A} \citep{Thoene2013}, and in imaging of the hosts (\object{GRB 051022}; \citealt{Castro-Tirado2007,Rol2007}, and \object{GRB060814}; \citealt{Perley2013}). Similarly, interacting systems have also been detected in the host systems of superluminous supernovae \citep{Chen2017,Perley2016c}.
However, the detection of mergers in GRB and SN hosts could be coincidental, especially at high redshift where mergers are much more frequent (e.g. \citealt{Callegari2009}). Both simulations and observations show that at $z\approx3$ more than one in ten galaxies of stellar mass $>10^9~ M_{\sun}$ have experienced a major merger in the preceding 1 Gyr \citep{Rodriguez-Gomez2015}. It remains to be clarified whether such GRBs are actually a result of the merger, or simply coincident. %While the SFR of $\sim 50 M_{\sun}/$year measured for the host of GRB 080810 is not unusual for a galaxy of $10^{10} M_{\sun}$ at $z=3.35$, the low densities of neutral gas suggest that there is little fuel for this star formation, and we speculate that the interaction may indeed have had a positive influence on the SFR and thus the chances of producing a GRB.

\subsection{Inflows and outflows\label{subsec:inout}}

The energetic, turbulent medium present in system A, the strong ionisation, and the mixing of different gaseous phases all point towards inhomogeneous motion throughout the ISM and CGM of the galaxy system forming the host of GRB 080810, on small and large scales. While there is already evidence for the presence of large scale outflows in GRB host galaxies  (e.g. \citealt{Schaefer2003,Chen2009}), system C is one of the most convincing yet. It has been strongly enriched with metals, and its ionisation and excitation properties suggest an origin in the star-forming regions of its galaxy, near where the GRB occurred. In contrast to system A which is highly-ionised and likely representative of a hot, wind-enriched galaxy halo, system C appears not yet to have reached its galaxy's CGM. Instead, it has maintained similar conditions to those where it was launched. 
 
A likely scenario for the geometry of the entire system is presented in Fig \ref{fig:scenario}. The locations of systems A, B, and C are fairly well constrained by their metal absorption. System D is harder to place, as there is little to be learned from its negligible metal content. It is highly improbable that it is an outflow similar to C but with a higher velocity, or indeed that it has originated inside the galaxy at all. Instead we believe it to trace the metal-poor gas that acts as a replenishment mechanism of galaxies throughout the universe. It has indeed been suggested that GRB hosts in general are fuelled by recent metal-poor gas inflow \citep{Michalowski2015,Michalowski2016}. The $-320$ \kms~blueshift implies that it is not accreting onto galaxy B --- if B hosted the GRB then D must be in front of it, moving away at that velocity. The degeneracy between local velocities and cosmological redshift allows for a range of locations of system D. One possibility is that it is located between the two galaxies and is currently falling towards galaxy A, or has fallen from behind galaxy B and is being slowed by its gravity. Another is that it has fallen from a large distance into the potential well of the complex, and is now in the foreground having initially passed both galaxies without being accreted, having joined in a disk-like structure (e.g. \citealt{Bouche2013}).
For any of these scenarios, system D represents a rare detection of such a metal-poor inflow in a GRB galaxy, or indeed in absorption at such a small projected distance to any galaxy. Systems with positive velocities relative to the gas local to GRBs, and thus inferred to be falling in towards the star-forming regions of their hosts, have been detected previously (e.g. \citealt{Prochaska2008a}), but these have tended to be richer in metals than the gas observed here, and may represent recycled gas that has previously been expelled from the galaxy.
Ly-$\alpha$ emitters are thought to trace accretion in large halos around quasars \citep{Cantalupo2014,Hennawi2015}, and have been detected at impact parameters of $\geq 70$ kpc from an LLS \citep{Fumagalli2016}. If system D is indeed bound to the host system, then it could be much closer to the galaxies than that.
Finally, as discussed when defining the host complex in Section \ref{subsec:host_lines}, it could in fact be a foreground system located $\sim$1.5 Mpc from the interacting complex. While this would exclude it from being a direct accretion flow onto the complex, it represents at the very least a reservoir of cool, metal-poor gas residing in the IGM near the dark matter halo, available for star-formation at some future epoch. 

\section{Summary and Conclusions\label{sec:conc}}
The rapid follow-up observations of the extremely bright \object{GRB 080810} with a high resolution spectrograph at a 10-m class telescope have provided us with an extremely rich window into the Universe at high redshift. Populated by numerous absorption systems from redshifts of upward of 2.3, the line of sight to the afterglow reveals a myriad of absorbing clouds in the IGM, sporadic denser systems with associated metal absorption, and a complex, extreme environment playing host to the GRB itself, dominated by an interacting, star-forming, Lyman-break galaxy system at $z\approx3.35$. 

In this paper, we have examined the properties of each of the four main systems at this redshift based on their absorption properties, through which we have provided one of the most detailed analyses of the multiple component nature of a galaxy at high redshift. The emerging picture is one of a complex environment, the standout results of which are summarised below.\\

($a$) \object{GRB 080810} exploded in the more distant partner of an interacting system. This component is hidden behind the compact foreground companion (which has the brightest Ly-$\alpha$ emission), and in total the system is forming stars at a rate of $\sim 10^2 M_{\sun} $yr$^{-1}$. The two systems are closing with a relative velocity of $\sim700$ \kms. 

($b$) The column density of neutral hydrogen measured throughout the entire complex is very low: even the strongest absorption is limited to $\log N_{\ion{H}{I}}^{\mathrm{B}} < 18.35$, one of the lowest ever measured in a GRB host system, despite the LOS to the burst passing within $\lesssim 5$ kiloparsecs of the centres of both galaxies. These are unprecedented scales for such a low level of neutral absorption.

($c$) Near the GRB site there is more neutral gas. The sight line through the interacting galaxy reveals a similar metal content but in a much higher ionisation state, representative of gas which has been expelled from the ISM into the hot, turbulent halo. The implication that this is the CGM at a galactic radius of $\lesssim 5$ kpc is remarkable. 

($d$) The mechanism by which such metals are expelled into the CGM is likely seen in the form of an outflowing cloud in the GRB host, which has a velocity of $\sim -200$ \kms ~relative to the denser gas nearest to the GRB.

($e$) The ongoing, intense star formation is likely fueled by metal-poor, inflowing gas. A candidate counterpart for this gas is seen in the spectrum, containing only marginal detections of heavy elements. Due to the degeneracy between cosmological redshift and local peculiar velocity, we cannot confirm that this gas is spatially coincident with the host galaxies. If it indeed belongs to the host complex, then it is one of the first detections to date of metal-poor gas in such close vicinity to star-forming galaxies.

\begin{acknowledgements}
We thank the referee for their detailed and helpful comments in improving this manuscript. We are grateful for the support of the GROND GRB team. We thank in particular Julia Bodensteiner, Ting-Wan Chen, and Sandra Savaglio for interesting and useful discussions. Some of the data presented in this manuscript were obtained at the W. M. Keck Observatory, which is operated as a scientific partnership among the California Institute of Technology, the University of California, and NASA; the Observatory was made possible by the generous financial support of the W. M. Keck Foundation. We wish to extend special thanks to those of Hawaiian ancestry on whose sacred mountain those from the astronomical community are privileged to be guests. We thank C. Ledoux and P. Vreeswijk for providing a compilation of oscillator strengths. PW, PS, TK, and R.M.Y acknowledge support through the Sofja Kovalevskaja Award to PS from the Alexander von Humboldt Foundation of Germany. AdUP acknowledges support from the Spanish research project AYA 2014-58381-P, from a Ram\'on y Cajal fellowship, and from a 2016 BBVA Foundation Grant for Researchers and Cultural Creators. We recognise the extensive use of the \texttt{NumPy}, \texttt{SciPy} \citep{VanderWalt2011}, and \texttt{matplotlib} \citep{Hunter2007} packages throughout the preparation of this manuscript.
\end{acknowledgements}
%-------------------------------------------------------------------
\bibliographystyle{aa} % style aa.bst
\bibliography{all_papers} % your references Yourfile.bib

\begin{thebibliography}{126}
\expandafter\ifx\csname natexlab\endcsname\relax\def\natexlab#1{#1}\fi

\bibitem[{Aguirre {et~al.}(2003)Aguirre, Schaye, Kim, Theuns, Rauch, \&
  Sargent}]{Aguirre2003}
Aguirre, A., Schaye, J., Kim, T.-S., {et~al.} 2003, ApJ, 602, 38

\bibitem[{Asplund {et~al.}(2009)Asplund, Grevesse, Sauval, \&
  Scott}]{Asplund2009}
Asplund, M., Grevesse, N., Sauval, A.~J., \& Scott, P. 2009, Astrophys. Space
  Sci., 47, 481

\bibitem[{Barthelmy {et~al.}(2005)Barthelmy, Barbier, Cummings, Fenimore,
  Gehrels, Hullinger, Krimm, Markwardt, Palmer, Parsons, Sato, Suzuki,
  Takahashi, Tashiro, \& Tueller}]{Barthelmy2005}
Barthelmy, S.~D., Barbier, L.~M., Cummings, J.~R., {et~al.} 2005, Sp. Sci. Rev.
  Vol. 120, Issue 3-4, pp. 143-164, 120, 143

\bibitem[{Bell {et~al.}(2006)Bell, Phleps, Somerville, Wolf, Borch, \&
  Meisenheimer}]{Bell2006}
Bell, E.~F., Phleps, S., Somerville, R.~S., {et~al.} 2006, ApJ, 652, 270

\bibitem[{Bernstein {et~al.}(2015)Bernstein, Burles, \&
  Prochaska}]{Bernstein2015}
Bernstein, R.~A., Burles, S.~M., \& Prochaska, J.~X. 2015, Publ. Astron. Soc.
  Pacific, 127, 911

\bibitem[{Bouch{\'{e}} {et~al.}(2013)Bouch{\'{e}}, Murphy, Kacprzak,
  P{\'{e}}roux, Contini, Martin, \& Dessauges-Zavadsky}]{Bouche2013}
Bouch{\'{e}}, N., Murphy, M.~T., Kacprzak, G.~G., {et~al.} 2013, Science, 341,
  50

\bibitem[{Bournaud(2011)}]{Bournaud2011}
Bournaud, F. 2011, EAS Publ. Ser., 51, 107

\bibitem[{Burrows {et~al.}(2005)Burrows, Hill, Nousek, Kennea, Wells, Osborne,
  Abbey, Beardmore, Mukerjee, Short, Chincarini, Campana, Citterio, Moretti,
  Pagani, Tagliaferri, Giommi, Capalbi, Tamburelli, Angelini, Cusumano,
  Br{\"{a}}uninger, Burkert, \& Hartner}]{Burrows2005}
Burrows, D.~N., Hill, J.~E., Nousek, J.~a., {et~al.} 2005, Space Sci. Rev.,
  120, 165

\bibitem[{Callegari {et~al.}(2009)Callegari, Mayer, Kazantzidis, Colpi,
  Governato, Quinn, \& Wadsley}]{Callegari2009}
Callegari, S., Mayer, L., Kazantzidis, S., {et~al.} 2009, ApJL, 696, L89

\bibitem[{Cano {et~al.}(2017)Cano, Wang, Dai, \& Wu}]{Cano2017}
Cano, Z., Wang, S.-Q., Dai, Z.-G., \& Wu, X.-F. 2017, in Adv. Astron. GRB
  Swift/Fermi Era Beyond, Accepted Review in

\bibitem[{Cantalupo {et~al.}(2014)Cantalupo, Arrigoni-Battaia, Prochaska,
  Hennawi, \& Madau}]{Cantalupo2014}
Cantalupo, S., Arrigoni-Battaia, F., Prochaska, J.~X., Hennawi, J.~F., \&
  Madau, P. 2014, Nature, 506, 63

\bibitem[{Castro-Tirado {et~al.}(2007)Castro-Tirado, Bremer, McBreen,
  Gorosabel, Guziy, Delgado, Bihain, Fakthullin, Pandey, Jelinek, Postigo,
  Sokolov, Misra, Sagar, Bama, Kamble, Anupama, Licandro, Aceituno, Neri,
  Aceituno, \& Neri}]{Castro-Tirado2007}
Castro-Tirado, A.~J., Bremer, M., McBreen, S., {et~al.} 2007, A{\&}A, 475, 101

\bibitem[{Cepa {et~al.}(2000)Cepa, Aguiar-Gonzalez, Gonzalez-Escalera,
  Gonzalez-Serrano, Joven-Alvarez, Cano, Rasilla, Rodriguez-Ramos, Gonzalez,
  {Cobos Duenas}, Sanchez, Tejada, Bland-Hawthorn, Militello, \&
  Rosa}]{Cepa2000}
Cepa, J., Aguiar-Gonzalez, M., Gonzalez-Escalera, V., {et~al.} 2000, in Soc.
  Photo-Optical Instrum. Eng. Conf. Ser., ed. M.~Iye \& A.~F.~M. Moorwood, Vol.
  4008, 623

\bibitem[{Chen(2011)}]{Chen2011}
Chen, H.-W. 2011, arXiv: 1110.0487

\bibitem[{Chen {et~al.}(2009)Chen, Perley, Pollack, Prochaska, Bloom,
  Dessauges-Zavadsky, Pettini, Lopez, {Dall 'aglio}, \& Becker}]{Chen2009}
Chen, H.-W., Perley, D.~A., Pollack, L.~K., {et~al.} 2009, ApJ, 691, 152

\bibitem[{Chen {et~al.}(2017)Chen, Nicholl, Smartt, Mazzali, Yates, Moriya,
  Inserra, Langer, Kruehler, Pan, Kotak, Galbany, Schady, Wiseman, Greiner,
  Schulze, Man, Jerkstrand, Smith, Dennefeld, Baltay, Bolmer, Kankare, Knust,
  Maguire, Rabinowitz, Rostami, Sullivan, \& Young}]{Chen2017}
Chen, T.~W., Nicholl, M., Smartt, S.~J., {et~al.} 2017, A{\&}A, 602, A9

\bibitem[{Chisholm {et~al.}(2016)Chisholm, Tremonti, Leitherer, Chen, \&
  Wofford}]{Chisholm2016}
Chisholm, J., Tremonti, C.~A., Leitherer, C., Chen, Y., \& Wofford, A. 2016,
  MNRAS, 457, 3133

\bibitem[{Christensen {et~al.}(2011)Christensen, Fynbo, Prochaska, Th{\"{o}}ne,
  {de Ugarte Postigo}, \& Jakobsson}]{Christensen2011}
Christensen, L., Fynbo, J. P.~U., Prochaska, J.~X., {et~al.} 2011, ApJ, 727, 73

\bibitem[{Crighton {et~al.}(2013)Crighton, Hennawi, \&
  Prochaska}]{Crighton2013}
Crighton, N. H.~M., Hennawi, J.~F., \& Prochaska, J.~X. 2013, ApJL, 776, 18

\bibitem[{Cucchiara {et~al.}(2015)Cucchiara, Fumagalli, Rafelski, Kocevski,
  Prochaska, Cooke, \& Becker}]{Cucchiara2015}
Cucchiara, A., Fumagalli, M., Rafelski, M., {et~al.} 2015, ApJ, 804, 51

\bibitem[{{Dalla Vecchia} \& Schaye(2008)}]{DallaVecchia2008}
{Dalla Vecchia}, C. \& Schaye, J. 2008, MNRAS, 387, 1431

\bibitem[{{Dalla Vecchia} \& Schaye(2012)}]{DallaVecchia2012}
{Dalla Vecchia}, C. \& Schaye, J. 2012, MNRAS, 426, 140

\bibitem[{{De Cia} {et~al.}(2016){De Cia}, Ledoux, Mattsson, Petitjean,
  Srianand, Gavignaud, \& Jenkins}]{DeCia2016}
{De Cia}, A., Ledoux, C., Mattsson, L., {et~al.} 2016, A{\&}A, 97, 1

\bibitem[{{De Cia} {et~al.}(2013){De Cia}, Ledoux, Savaglio, Schady, \&
  Vreeswijk}]{DeCia2013}
{De Cia}, A., Ledoux, C., Savaglio, S., Schady, P., \& Vreeswijk, P.~M. 2013,
  A{\&}A, 560, A88

\bibitem[{{de Ugarte Postigo} {et~al.}(2012){de Ugarte Postigo}, Fynbo,
  Th{\"{o}}ne, Christensen, Gorosabel, Milvang-Jensen, Schulze, Jakobsson,
  Wiersema, Sanchez-Ramirez, Leloudas, Zafar, Malesani, \&
  Hjorth}]{DeUgartePostigo2012}
{de Ugarte Postigo}, A., Fynbo, J. P.~U., Th{\"{o}}ne, C.~C., {et~al.} 2012,
  A{\&}A, 548, A11

\bibitem[{Dekel {et~al.}(2006)Dekel, Birnboim, K., M., F., J., J., L., A., A.,
  B., M., Y., A., W., J., G., G., V., S., A., C., U., J., S., A., G., K., M.,
  F., F., F., J., J., A., Y., R., R., R., L., S., S., M., A., C., C., L., S.,
  J., G., A., A., A., A., M., J., A., A., A., M., M., A., C., M., R., Z., F.,
  S., A., F., G., D., A., M., V., V., J., O., A., C., P., S., H., H., H., C.,
  G., J., J., N., D., F., F., K., F., X., J., M., E., J., E., J., K., A., I.,
  S., D., A., M., M., M., R., \& X.}]{Dekel2006}
Dekel, A., Birnboim, Y., K., A., {et~al.} 2006, MNRAS, 368, 2

\bibitem[{D'Elia {et~al.}(2011)D'Elia, Campana, Covino, D'Avanzo, Piranomonte,
  \& Tagliaferri}]{DElia2011}
D'Elia, V., Campana, S., Covino, S., {et~al.} 2011, MNRAS, 418, 680

\bibitem[{D'Elia {et~al.}(2014)D'Elia, Fynbo, Goldoni, Covino, {de Ugarte
  Postigo}, Ledoux, Calura, Gorosabel, Malesani, Matteucci,
  S{\'{a}}nchez-Ram{\'{i}}rez, Savaglio, Castro-Tirado, Hartoog, Kaper,
  Mu{\~{n}}oz-Darias, Pian, Piranomonte, Tagliaferri, Tanvir, Vergani, Watson,
  \& Xu}]{DElia2014}
D'Elia, V., Fynbo, J. P.~U., Goldoni, P., {et~al.} 2014, A{\&}A, 564, A38

\bibitem[{Dessauges-Zavadsky {et~al.}(2006)Dessauges-Zavadsky, Prochaska,
  D'Odorico, Calura, \& Matteucci}]{Dessauges-Zavadsky2006}
Dessauges-Zavadsky, M., Prochaska, J.~X., D'Odorico, S., Calura, F., \&
  Matteucci, F. 2006, A{\&}A, 445, 93

\bibitem[{Ferrero {et~al.}(2009)Ferrero, Klose, Kann, Savaglio, Schulze,
  Palazzi, Maiorano, B{\"{o}}hm, Grupe, Oates, S{\'{a}}nchez, Amati, Greiner,
  Hjorth, Malesani, Barthelmy, Gorosabel, Masetti, \& Roth}]{Ferrero2009}
Ferrero, P., Klose, S., Kann, D.~A., {et~al.} 2009, A{\&}A, 497, 729

\bibitem[{Finley {et~al.}(2017)Finley, Bouch{\'{e}}, Contini, Epinat, Bacon,
  Brinchmann, Cantalupo, Erroz-Ferrer, Marino, Maseda, Richard, Verhamme,
  Weilbacher, Wendt, \& Wisotzki}]{Finley2017}
Finley, H., Bouch{\'{e}}, N., Contini, T., {et~al.} 2017, arXiv: 1701.07843

\bibitem[{Fox {et~al.}(2008)Fox, Ledoux, Vreeswijk, Smette, \&
  Jaunsen}]{Fox2008}
Fox, A.~J., Ledoux, C., Vreeswijk, P.~M., Smette, A., \& Jaunsen, A.~O. 2008,
  A{\&}A, 491, 189

\bibitem[{Fraternali \& Filippo(2016)}]{Fraternali2016}
Fraternali, F. \& Filippo. 2016, in Gas Accretion onto Galaxies, Astrophys. Sp.
  Sci. Libr. eds. A. J. Fox R. Dave, 2017, Publ. by Springer

\bibitem[{Friis {et~al.}(2015)Friis, {De Cia}, Kr{\"{u}}hler, Fynbo, Ledoux,
  Vreeswijk, Watson, Malesani, Gorosabel, Starling, Jakobsson, Varela,
  Wiersema, Drachmann, Trotter, Th{\"{o}}ne, {De Ugarte Postigo}, D'Elia,
  Elliott, Maturi, Goldoni, Greiner, Haislip, Kaper, Knust, LaCluyze,
  Milvang-Jensen, Reichart, Schulze, Sudilovsky, Tanvir, \&
  Vergani}]{Friis2015}
Friis, M., {De Cia}, A., Kr{\"{u}}hler, T., {et~al.} 2015, MNRAS, 451, 167

\bibitem[{Fumagalli {et~al.}(2016)Fumagalli, Cantalupo, Dekel, Morris, O'Meara,
  Prochaska, \& Theuns}]{Fumagalli2016}
Fumagalli, M., Cantalupo, S., Dekel, A., {et~al.} 2016, MNRAS, 462, 1978

\bibitem[{Fumagalli {et~al.}(2011{\natexlab{a}})Fumagalli, O'Meara, \&
  Prochaska}]{Fumagalli2011a}
Fumagalli, M., O'Meara, J.~M., \& Prochaska, J.~X. 2011{\natexlab{a}}, Science,
  334, 1245

\bibitem[{Fumagalli {et~al.}(2013)Fumagalli, O'Meara, Prochaska, \&
  Worseck}]{Fumagalli2013}
Fumagalli, M., O'Meara, J.~M., Prochaska, J.~X., \& Worseck, G. 2013, ApJ, 775,
  78

\bibitem[{Fumagalli {et~al.}(2011{\natexlab{b}})Fumagalli, Prochaska, Kasen,
  Dekel, Ceverino, \& Primack}]{Fumagalli2011b}
Fumagalli, M., Prochaska, J.~X., Kasen, D., {et~al.} 2011{\natexlab{b}}, MNRAS,
  418, 1796

\bibitem[{Fynbo {et~al.}(2009)Fynbo, Jakobsson, Prochaska, Malesani, Ledoux,
  {de Ugarte Postigo}, Nardini, Vreeswijk, Wiersema, Hjorth, Sollerman, Chen,
  Th{\"{o}}ne, Bj{\"{o}}rnsson, Bloom, Castro-Tirado, Christensen, {De Cia},
  Fruchter, Gorosabel, Graham, Jaunsen, Jensen, Kann, Kouveliotou, Levan,
  Maund, Masetti, Milvang-Jensen, Palazzi, Perley, Pian, Rol, Schady, Starling,
  Tanvir, Watson, Xu, Augusteijn, Grundahl, Telting, \& Quirion}]{Fynbo2009}
Fynbo, J. P.~U., Jakobsson, P., Prochaska, J.~X., {et~al.} 2009, ApJS, 185, 526

\bibitem[{Fynbo {et~al.}(2006)Fynbo, Starling, Ledoux, Wiersema, Th{\"{o}}ne,
  Sollerman, Jakobsson, Hjorth, Watson, Vreeswijk, M{\o}ller, Rol, Gorosabel,
  N{\"{a}}r{\"{a}}nen, Wijers, Bj{\"{o}}rnsson, Cer{\'{o}}n, Curran, Hartmann,
  Holland, Jensen, Levan, Limousin, Kouveliotou, Nelemans, Pedersen, Priddey,
  \& Tanvir}]{Fynbo2006}
Fynbo, J. P.~U., Starling, R. L.~C., Ledoux, C., {et~al.} 2006, A{\&}A, 451,
  L47

\bibitem[{Galama {et~al.}(1998)Galama, Vreeswijk, van Paradijs, Kouveliotou,
  Augusteijn, Hainaut, Patat, Boehnhardt, Brewer, Doublier, Gonzalez, Lidman,
  Leibundgut, Heise, Zand, Groot, Strom, Mazzali, Iwamoto, Nomoto, Umeda,
  Nakamura, Koshut, Kippen, Robinson, de~Wildt, Wijers, Tanvir, Greiner, Pian,
  Palazzi, Frontera, Masetti, Nicastro, Malozzi, Feroci, Costa, Piro, Peterson,
  Tinney, Boyle, Cannon, Stathakis, Begam, \& Ianna}]{Galama1998}
Galama, T.~J., Vreeswijk, P.~M., van Paradijs, J., {et~al.} 1998, Nature, 395,
  670

\bibitem[{Gehrels {et~al.}(2004)Gehrels, Chincarini, Giommi, Mason, Nousek,
  Wells, White, Barthelmy, Burrows, Cominsky, Hurley, Marshall, Me?sza?ros,
  Roming, Angelini, Barbier, Belloni, Campana, Caraveo, Chester, Citterio,
  Cline, Cropper, Cummings, Dean, Feigelson, Fenimore, Frail, Fruchter,
  Garmire, Gendreau, Ghisellini, Greiner, Hill, Hunsberger, Krimm, Kulkarni,
  Kumar, Lebrun, Lloyd-Ronning, Markwardt, Mattson, Mushotzky, Norris, Osborne,
  Paczynski, Palmer, Park, Parsons, Paul, Rees, Reynolds, Rhoads, Sasseen,
  Schaefer, Short, Smale, Smith, Stella, Tagliaferri, Takahashi, Tashiro,
  Townsley, Tueller, Turner, Vietri, Voges, Ward, Willingale, Zerbi, \&
  Zhang}]{Gehrels2004}
Gehrels, N., Chincarini, G., Giommi, P., {et~al.} 2004, ApJ, 611, 1005

\bibitem[{Glidden {et~al.}(2016)Glidden, Cooper, Cooksey, Simcoe, O'Meara, \&
  D.}]{Glidden2016}
Glidden, A., Cooper, T.~J., Cooksey, K.~L., {et~al.} 2016, ApJ, 833, 270

\bibitem[{Greiner {et~al.}(2015{\natexlab{a}})Greiner, Fox, Schady,
  Kr{\"{u}}hler, Trenti, Cikota, Bolmer, Elliott, Delvaux, Perna, Afonso, Kann,
  Klose, Savaglio, Schmidl, Schweyer, Tanga, \& Varela}]{Greiner2015a}
Greiner, J., Fox, D.~B., Schady, P., {et~al.} 2015{\natexlab{a}}, ApJ, 809, 76

\bibitem[{Greiner {et~al.}(2015{\natexlab{b}})Greiner, Mazzali, Kann,
  Kr{\"{u}}hler, Pian, Prentice, E., Rossi, Klose, Taubenberger, Knust, Afonso,
  Ashall, Bolmer, Delvaux, Diehl, Elliott, Filgas, Fynbo, Graham, Guelbenzu,
  Kobayashi, Leloudas, Savaglio, Schady, Schmid, Schweyer, Sudilovsky, Tanga,
  Updike, van Eerten, \& Varela}]{Greiner2015}
Greiner, J., Mazzali, P.~a., Kann, D.~A., {et~al.} 2015{\natexlab{b}}, Nature,
  523, 189

\bibitem[{Grimes {et~al.}(2009)Grimes, Heckman, Aloisi, Calzetti, Leitherer,
  Martin, Meurer, Sembach, \& Strickland}]{Grimes2009}
Grimes, J.~P., Heckman, T., Aloisi, A., {et~al.} 2009, ApJS, 181, 272

\bibitem[{Hartoog {et~al.}(2015)Hartoog, Malesani, Fynbo, Goto, Kr{\"{u}}hler,
  Vreeswijk, Cia, \& Xu}]{Hartoog2015}
Hartoog, O.~E., Malesani, D., Fynbo, J. P.~U., {et~al.} 2015, A{\&}A, 139, 1

\bibitem[{Heckman {et~al.}(2015)Heckman, Alexandroff, Borthakur, Overzier, \&
  Leitherer}]{Heckman2015}
Heckman, T.~M., Alexandroff, R.~M., Borthakur, S., Overzier, R., \& Leitherer,
  C. 2015, ApJ, 809, 147

\bibitem[{Hennawi {et~al.}(2015)Hennawi, Prochaska, Cantalupo, \&
  Arrigoni-Battaia}]{Hennawi2015}
Hennawi, J.~F., Prochaska, J.~X., Cantalupo, S., \& Arrigoni-Battaia, F. 2015,
  Science, 348, 779

\bibitem[{Henriques {et~al.}(2015)Henriques, White, Thomas, Angulo, Guo,
  Lemson, Springel, \& Overzier}]{Henriques2015}
Henriques, B., White, S., Thomas, P., {et~al.} 2015, MNRAS, 451, 2663

\bibitem[{Hjorth {et~al.}(2003)Hjorth, Sollerman, M{\o}ller, Fynbo, Woosley,
  Kouveliotou, Tanvir, Greiner, Andersen, Castro-Tirado, Cer{\'{o}}n, Fruchter,
  Gorosabel, Jakobsson, Kaper, Klose, Masetti, Pedersen, Pedersen, Pian,
  Palazzi, Rhoads, Rol, van~den Heuvel, Vreeswijk, Watson, \&
  Wijers}]{Hjorth2003}
Hjorth, J., Sollerman, J., M{\o}ller, P., {et~al.} 2003, Nature, 423, 847

\bibitem[{Hunter(2007)}]{Hunter2007}
Hunter, J.~D. 2007, Comput. Sci. Eng., 9, 90

\bibitem[{Jenkins(2009)}]{Jenkins2009}
Jenkins, E.~B. 2009, ApJ, 700, 1299

\bibitem[{Jorgenson {et~al.}(2013)Jorgenson, Murphy, \&
  Thompson}]{Jorgenson2013}
Jorgenson, R.~A., Murphy, M.~T., \& Thompson, R. 2013, MNRAS, 435, 482

\bibitem[{Kann {et~al.}(2010)Kann, Klose, Zhang, Malesani, Nakar, Pozanenko,
  Wilson, Butler, Jakobsson, Schulze, Andreev, Antonelli, Bikmaev, Biryukov,
  B{\"{o}}ttcher, Burenin, {Castro Cer{\'{o}}n}, Castro-Tirado, Chincarini,
  Cobb, Covino, D'Avanzo, D'Elia, {Della Valle}, {de Ugarte Postigo}, Efimov,
  Ferrero, Fugazza, Fynbo, G{\aa}lfalk, Grundahl, Gorosabel, Gupta, Guziy,
  Hafizov, Hjorth, Holhjem, Ibrahimov, Im, Israel, Jeĺinek, Jensen, Karimov,
  Khamitov, Kiziloǧlu, Klunko, Kub{\'{a}}nek, Kutyrev, Laursen, Levan,
  Mannucci, Martin, Mescheryakov, Mirabal, Norris, Ovaldsen, Paraficz,
  Pavlenko, Piranomonte, Rossi, Rumyantsev, Salinas, Sergeev, Sharapov,
  Sollerman, Stecklum, Stella, Tagliaferri, Tanvir, Telting, Testa, Updike,
  Volnova, Watson, Wiersema, \& Xu}]{Kann2010}
Kann, D.~A., Klose, S., Zhang, B., {et~al.} 2010, ApJ, 720, 1513

\bibitem[{Kr{\"{u}}hler {et~al.}(2013)Kr{\"{u}}hler, Ledoux, Fynbo, Vreeswijk,
  Schmidl, Malesani, Christensen, {De Cia}, Hjorth, Jakobsson, Kann, Kaper,
  Vergani, Afonso, Covino, {de Ugarte Postigo}, D'Elia, Filgas, Goldoni,
  Greiner, Hartoog, Milvang-Jensen, Nardini, Piranomonte, Rossi,
  Sanchez-Ramirez, Schady, Schulze, Sudilovsky, Tanvir, Tagliaferri, Watson,
  Wiersema, Wijers, \& Xu}]{Kruehler2013}
Kr{\"{u}}hler, T., Ledoux, C., Fynbo, J. P.~U., {et~al.} 2013, A{\&}A, 557, A18

\bibitem[{Larson \& B.(1974)}]{Larson1974}
Larson, R.~B. \& B., R. 1974, MNRAS, 169, 229

\bibitem[{Ledoux {et~al.}(2009)Ledoux, Vreeswijk, Smette, Fox, Petitjean,
  Ellison, Fynbo, \& Savaglio}]{Ledoux2009}
Ledoux, C., Vreeswijk, P.~M., Smette, A., {et~al.} 2009, A{\&}A, 506, 661

\bibitem[{Lehner {et~al.}(2013)Lehner, Howk, Tripp, Tumlinson, Prochaska,
  O'Meara, Thom, Werk, Fox, \& Ribaudo}]{Lehner2013}
Lehner, N., Howk, J.~C., Tripp, T.~M., {et~al.} 2013, ApJ, 770, 138

\bibitem[{Lehner \& Nicolas(2016)}]{Lehner2016}
Lehner, N. \& Nicolas. 2016, in Gas Accretion onto Galaxies, Astrophys. Sp.
  Sci. Libr. eds. A. J. Fox R. Dave, 2017, Publ. by Springer

\bibitem[{Lehner {et~al.}(2016)Lehner, O'Meara, Howk, Prochaska, \&
  Fumagalli}]{Lehner2016a}
Lehner, N., O'Meara, J.~M., Howk, J.~C., Prochaska, J.~X., \& Fumagalli, M.
  2016, ApJ, 833, 283

\bibitem[{Lopez-Sanchez \& Esteban(2009)}]{Lopez-Sanchez2009}
Lopez-Sanchez, A.~R. \& Esteban, C. 2009, A{\&}A, 508, 615

\bibitem[{Martin(1999)}]{Martin1999}
Martin, C.~L. 1999, ApJ, 513, 156

\bibitem[{Martin(2005)}]{Martin2005}
Martin, C.~L. 2005, ApJ, 621, 227

\bibitem[{Mathews \& Baker(1971)}]{Mathews1971}
Mathews, W.~G. \& Baker, J.~C. 1971, ApJ, 170, 241

\bibitem[{McLean {et~al.}(2012)McLean, Steidel, Epps, Konidaris, Matthews,
  Adkins, Aliado, Brims, Canfield, Cromer, Fucik, Kulas, Mace, Magnone,
  Rodriguez, Rudie, Trainor, Wang, Weber, \& Weiss}]{McLean+2012}
McLean, I., Steidel, C., Epps, H., {et~al.} 2012, in Soc. Photo-Optical
  Instrum. Eng. Conf. Ser., Vol. 8446

\bibitem[{Micha{\l}owski {et~al.}(2016)Micha{\l}owski, Ceron, Wardlow, Karska,
  Messias, van~der Werf, Hunt, Baes, Castro-Tirado, Gentile, Hjorth, Floc'h,
  Martinez, Guelbenzu, Rasmussen, Rizzo, Rossi, Sanchez-Portal, Schady,
  Sollerman, \& Xu}]{Michalowski2016}
Micha{\l}owski, M.~J., Ceron, J. M.~C., Wardlow, J.~L., {et~al.} 2016, A{\&}A,
  595, A72

\bibitem[{Micha{\l}owski {et~al.}(2015)Micha{\l}owski, Gentile, Hjorth,
  Krumholz, Tanvir, Kamphuis, Burlon, Baes, Basa, Berta, {Castro Cer{\'{o}}n},
  Crosby, D'Elia, Elliott, Greiner, Hunt, Klose, Koprowski, {Le Floc'h},
  Malesani, Murphy, {Nicuesa Guelbenzu}, Palazzi, Rasmussen, Rossi, Savaglio,
  Schady, Sollerman, {de Ugarte Postigo}, Watson, van~der Werf, Vergani, \&
  Xu}]{Michalowski2015}
Micha{\l}owski, M.~J., Gentile, G., Hjorth, J., {et~al.} 2015, A{\&}A, 582, A78

\bibitem[{Oke {et~al.}(1995)Oke, Cohen, Carr, Cromer, Dingizian, Harris,
  Labrecque, Lucinio, Schaal, Epps, \& Miller}]{Oke+1995}
Oke, J., Cohen, J., Carr, M., {et~al.} 1995, PASP, 107, 375

\bibitem[{Page {et~al.}(2008)Page, Barthelmy, Burrows, Evans, Guidorzi,
  Holland, Kennea, Kuin, Mao, Marshall, Palmer, Stamatikos, Starling, \&
  Ukwatta}]{Page2008}
Page, K.~L., Barthelmy, S.~D., Burrows, D.~N., {et~al.} 2008, GRB Coord. Netw.,
  8080

\bibitem[{Page {et~al.}(2009)Page, Willingale, Bissaldi, Postigo, Holland,
  McBreen, O'Brien, Osborne, Prochaska, Rol, Rykoff, Starling, Tanvir, van~der
  Horst, Wiersema, Zhang, Aceituno, Akerlof, Beardmore, Briggs, Burrows,
  Castro-Tirado, Connaughton, Evans, Fynbo, Gehrels, Guidorzi, Howard, Kennea,
  Kouveliotou, Pagani, Preece, Perley, Steele, Yuan, \& Yuan}]{Page2009}
Page, K.~L., Willingale, R., Bissaldi, E., {et~al.} 2009, MNRAS, 400, 134

\bibitem[{Perley {et~al.}(2013)Perley, Levan, Tanvir, Cenko, Bloom, Hjorth,
  Kruehler, Filippenko, Fruchter, Fynbo, Jakobsson, Kalirai, Milvang-Jensen,
  Morgan, Prochaska, \& Silverman}]{Perley2013}
Perley, D.~A., Levan, A.~J., Tanvir, N.~R., {et~al.} 2013, ApJ, 778, 128

\bibitem[{Perley {et~al.}(2012)Perley, Modjaz, Morgan, Cenko, Bloom, Butler,
  Filippenko, \& Miller}]{Perley2012}
Perley, D.~A., Modjaz, M., Morgan, A.~N., {et~al.} 2012, ApJ, 758

\bibitem[{Perley {et~al.}(2016{\natexlab{a}})Perley, Quimby, Yan, Vreeswijk,
  Cia, Lunnan, Gal-Yam, Yaron, Filippenko, Graham, Laher, \&
  Nugent}]{Perley2016c}
Perley, D.~A., Quimby, R.~M., Yan, L., {et~al.} 2016{\natexlab{a}}, ApJ, 830,
  13

\bibitem[{Perley {et~al.}(2016{\natexlab{b}})Perley, Tanvir, Hjorth, Laskar,
  Berger, Chary, Postigo, Fynbo, Kr{\"{u}}hler, Levan, Micha{\l}owski, \&
  Schulze}]{Perley2016b}
Perley, D.~A., Tanvir, N.~R., Hjorth, J., {et~al.} 2016{\natexlab{b}}, ApJ,
  817, 8

\bibitem[{Peroux {et~al.}(2007)Peroux, Dessauges-Zavadsky, D'Odorico, Kim, \&
  McMahon}]{Peroux2007}
Peroux, C., Dessauges-Zavadsky, M., D'Odorico, S., Kim, T.-S., \& McMahon,
  R.~G. 2007, MNRAS, 382, 177

\bibitem[{Pettini {et~al.}(2002)Pettini, Rix, Steidel, Adelberger, Hunt, \&
  Shapley}]{Pettini2002}
Pettini, M., Rix, S.~A., Steidel, C.~C., {et~al.} 2002, ApJ, 569, 742

\bibitem[{Phillips {et~al.}(1982)Phillips, Gondhalekar, \&
  Pettini}]{Phillips1982}
Phillips, A.~P., Gondhalekar, P.~M., \& Pettini, M. 1982, MNRAS, 200, 687

\bibitem[{Pieri {et~al.}(2013)Pieri, Mortonson, Frank, Crighton, Weinberg, Lee,
  Noterdaeme, Bailey, Busca, Ge, Kirkby, Lundgren, Mathur, Paris,
  Palanque-Delabrouille, Petitjean, Rich, Ross, Schneider, \& York}]{Pieri2013}
Pieri, M.~M., Mortonson, M.~J., Frank, S., {et~al.} 2013, MNRAS, 441, 1718

\bibitem[{{Planck Collaboration}(2014)}]{PlanckCollaboration2014}
{Planck Collaboration}. 2014, A{\&}A, 571, A16

\bibitem[{Prochaska {et~al.}(2006)Prochaska, Chen, \& Bloom}]{Prochaska2006b}
Prochaska, J.~X., Chen, H., \& Bloom, J.~S. 2006, ApJ, 648, 95

\bibitem[{Prochaska {et~al.}(2007{\natexlab{a}})Prochaska, Chen, Bloom,
  Dessauges-Zavadsky, O'Meara, Foley, Bernstein, Burles, Dupree, Falco, \&
  Thompson}]{Prochaska2007a}
Prochaska, J.~X., Chen, H., Bloom, J.~S., {et~al.} 2007{\natexlab{a}}, ApJS,
  168, 231

\bibitem[{Prochaska {et~al.}(2007{\natexlab{b}})Prochaska, Chen,
  Dessauges-Zavadsky, \& Bloom}]{Prochaska2007b}
Prochaska, J.~X., Chen, H., Dessauges-Zavadsky, M., \& Bloom, J.~S.
  2007{\natexlab{b}}, ApJ, 666, 267

\bibitem[{Prochaska {et~al.}(2008)Prochaska, Chen, Wolfe, Dessauges-Zavadsky,
  \& Bloom}]{Prochaska2008a}
Prochaska, J.~X., Chen, H., Wolfe, A.~M., Dessauges-Zavadsky, M., \& Bloom,
  J.~S. 2008, ApJ, 672, 59

\bibitem[{Prochaska {et~al.}(2015)Prochaska, O'Meara, Fumagalli, Bernstein, \&
  Burles}]{Prochaska2015}
Prochaska, J.~X., O'Meara, J.~M., Fumagalli, M., Bernstein, R.~A., \& Burles,
  S.~M. 2015, ApJS, 221, 2

\bibitem[{Prochaska {et~al.}(2017)Prochaska, Werk, Worseck, Tripp, Tumlinson,
  Burchett, Fox, Fumagalli, Lehner, Peeples, \& Tejos}]{Prochaska2017}
Prochaska, J.~X., Werk, J.~K., Worseck, G., {et~al.} 2017, ApJ, 837, 169

\bibitem[{Prochaska \& Wolfe(1998)}]{Prochaska1998}
Prochaska, J.~X. \& Wolfe, A.~M. 1998, ApJS, 121, 369

\bibitem[{Prochaska \& Wolfe(2002)}]{Prochaska2002a}
Prochaska, J.~X. \& Wolfe, A.~M. 2002, ApJ, 566, 68

\bibitem[{Prochter {et~al.}(2010)Prochter, Prochaska, O'Meara, Burles, \&
  Bernstein}]{Prochter2010}
Prochter, G.~E., Prochaska, J.~X., O'Meara, J.~M., Burles, S., \& Bernstein,
  R.~A. 2010, ApJ, 708, 1221

\bibitem[{Quiret {et~al.}(2016)Quiret, P{\'{e}}roux, Zafar, Kulkarni, Jenkins,
  Milliard, Rahmani, Popping, Rao, Turnshek, \& Monier}]{Quiret2016}
Quiret, S., P{\'{e}}roux, C., Zafar, T., {et~al.} 2016, MNRAS, 458, 4074

\bibitem[{Rafelski {et~al.}(2014)Rafelski, Neeleman, Fumagalli, Wolfe, \&
  Prochaska}]{Rafelski2014}
Rafelski, M., Neeleman, M., Fumagalli, M., Wolfe, A.~M., \& Prochaska, J.~X.
  2014, ApJL, 782, L29

\bibitem[{Rafelski {et~al.}(2012)Rafelski, Wolfe, Prochaska, Neeleman, \&
  Mendez}]{Rafelski2012}
Rafelski, M., Wolfe, A.~M., Prochaska, J.~X., Neeleman, M., \& Mendez, A.~J.
  2012, ApJ, 755, 89

\bibitem[{Rees \& Ostriker(1977)}]{Rees1977}
Rees, M.~J. \& Ostriker, J.~P. 1977, MNRAS, 179, 541

\bibitem[{Robaina {et~al.}(2009)Robaina, Bell, Skelton, McIntosh, Somerville,
  Zheng, Rix, Bacon, Balogh, Barazza, Barden, Boehm, Caldwell, Gallazzi, Gray,
  Haussler, Heymans, Jahnke, Jogee, van Kampen, Lane, Meisenheimer, Papovich,
  Peng, Sanchez, Skibba, Taylor, Wisotzki, \& Wolf}]{Robaina2009}
Robaina, A.~R., Bell, E.~F., Skelton, R.~E., {et~al.} 2009, ApJ, 704, 324

\bibitem[{Rodriguez-Gomez {et~al.}(2015)Rodriguez-Gomez, Genel, Vogelsberger,
  Sijacki, Pillepich, Sales, Torrey, Snyder, Nelson, Springel, Ma, \&
  Hernquist}]{Rodriguez-Gomez2015}
Rodriguez-Gomez, V., Genel, S., Vogelsberger, M., {et~al.} 2015, MNRAS, 449, 49

\bibitem[{Rol {et~al.}(2007)Rol, van~der Horst, Wiersema, Patel, Levan,
  Nysewander, Kouveliotou, Wijers, Tanvir, Reichart, Fruchter, Graham,
  Ovaldsen, Jaunsen, Jonker, van Ham, Hjorth, Starling, O'Brien, Fynbo,
  Burrows, \& Strom}]{Rol2007}
Rol, E., van~der Horst, A., Wiersema, K., {et~al.} 2007, ApJ, 669, 1098

\bibitem[{Roming {et~al.}(2005)Roming, Kennedy, Mason, Nousek, Ahr, Bingham,
  Broos, Carter, Hancock, Huckle, Hunsberger, Kawakami, Killough, Koch,
  McLelland, Smith, Smith, Soto, Boyd, Breeveld, Holland, Ivanushkina, Pryzby,
  Still, \& Stock}]{Roming2005}
Roming, P. W.~A., Kennedy, T.~E., Mason, K.~O., {et~al.} 2005, Space Sci. Rev.,
  120, 95

\bibitem[{Rubin {et~al.}(2012)Rubin, Prochaska, Koo, \& Phillips}]{Rubin2012}
Rubin, K. H.~R., Prochaska, J.~X., Koo, D.~C., \& Phillips, A.~C. 2012, ApJL,
  747, 26

\bibitem[{Rubin {et~al.}(2014)Rubin, Prochaska, Koo, Phillips, Martin, \&
  Winstrom}]{Rubin2014}
Rubin, K. H.~R., Prochaska, J.~X., Koo, D.~C., {et~al.} 2014, ApJ, 794, 156

\bibitem[{Sanders {et~al.}(1988)Sanders, Soifer, Elias, Madore, Matthews,
  Neugebauer, \& Scoville}]{Sanders1988}
Sanders, D.~B., Soifer, B.~T., Elias, J.~H., {et~al.} 1988, ApJ, 325, 74

\bibitem[{Savage \& Sembach(1996)}]{Savage1996}
Savage, B. D.~B. \& Sembach, K. R.~K. 1996, ARA{\&}A, 34, 279

\bibitem[{Savaglio {et~al.}(2003)Savaglio, Fall, \& Fiore}]{Savaglio2003}
Savaglio, S., Fall, S.~M., \& Fiore, F. 2003, ApJ, 585, 638

\bibitem[{Savaglio {et~al.}(2012)Savaglio, Rau, Greiner, Kr??hler, Mcbreen,
  Hartmann, Updike, Filgas, Klose, Afonso, Clemens, {K??pc?? Yolda??},
  {Olivares E.}, Sudilovsky, \& Szokoly}]{Savaglio2012}
Savaglio, S., Rau, A., Greiner, J., {et~al.} 2012, MNRAS, 420, 627

\bibitem[{Schady {et~al.}(2012)Schady, Dwelly, Page, Kr{\"{u}}hler, Greiner,
  Oates, {De Pasquale}, Nardini, Roming, Rossi, \& Still}]{Schady2012}
Schady, P., Dwelly, T., Page, M.~J., {et~al.} 2012, A{\&}A, 536, A15

\bibitem[{Schady {et~al.}(2011)Schady, Savaglio, Kr{\"{u}}hler, Greiner, \&
  Rau}]{Schady2011}
Schady, P., Savaglio, S., Kr{\"{u}}hler, T., Greiner, J., \& Rau, A. 2011,
  A{\&}A, 525, 1

\bibitem[{Schaefer {et~al.}(2003)Schaefer, Gerardy, Hoflich, Panaitescu,
  Quimby, Mader, Hill, Kumar, Wheeler, Eracleous, Sigurdsson, Meszaros, Zhang,
  Wang, Hessman, \& Petrosian}]{Schaefer2003}
Schaefer, B.~E., Gerardy, C.~L., Hoflich, P., {et~al.} 2003, ApJ, 588, 387

\bibitem[{Shapley {et~al.}(2003)Shapley, Steidel, Pettini, \&
  Adelberger}]{Shapley2003}
Shapley, A.~E., Steidel, C.~C., Pettini, M., \& Adelberger, K.~L. 2003, ApJ,
  588, 65

\bibitem[{Sparre {et~al.}(2014)Sparre, Hartoog, Kr{\"{u}}hler, Fynbo, Watson,
  Wiersema, D'Elia, Zafar, Afonso, Covino, {de Ugarte Postigo}, Flores,
  Goldoni, Greiner, Hjorth, Jakobsson, Kaper, Klose, Levan, Malesani,
  Milvang-Jensen, Nardini, Piranomonte, Sollerman, S{\'{a}}nchez-Ram{\'{i}}rez,
  Schulze, Tanvir, Vergani, \& Wijers}]{Sparre2014}
Sparre, M., Hartoog, O.~E., Kr{\"{u}}hler, T., {et~al.} 2014, ApJ, 785, 150

\bibitem[{Steidel {et~al.}(2010)Steidel, Erb, Shapley, Pettini, Reddy,
  Bogosavljevi{\'{c}}, Rudie, \& Rakic}]{Steidel2010}
Steidel, C.~C., Erb, D.~K., Shapley, A.~E., {et~al.} 2010, ApJ, 717, 289

\bibitem[{Th{\"{o}}ne {et~al.}(2013)Th{\"{o}}ne, Fynbo, Goldoni, {de Ugarte
  Postigo}, Campana, Vergani, Covino, Kr??hler, Kaper, Tanvir, Zafar, D'Elia,
  Gorosabel, Greiner, Groot, Hammer, Jakobsson, Klose, Levan, Milvang-Jensen,
  {Nicuesa Guelbenzu}, Palazzi, Piranomonte, Tagliaferri, Watson, Wiersema, \&
  Wijers}]{Thoene2013}
Th{\"{o}}ne, C.~C., Fynbo, J. P.~U., Goldoni, P., {et~al.} 2013, MNRAS, 428,
  3590

\bibitem[{van~der Walt {et~al.}(2011)van~der Walt, Colbert, \&
  Varoquaux}]{VanderWalt2011}
van~der Walt, S., Colbert, S.~C., \& Varoquaux, G. 2011, Comput. Sci. Eng., 13,
  22

\bibitem[{Viegas(1995)}]{Viegas1995}
Viegas, S.~M. 1995, MNRAS, 276, 268

\bibitem[{Vladilo(1998)}]{Vladilo1998}
Vladilo, G. 1998, ApJ, 493, 583

\bibitem[{Vladilo {et~al.}(2011)Vladilo, Abate, Yin, Cescutti, \&
  Matteucci}]{Vladilo2011}
Vladilo, G., Abate, C., Yin, J., Cescutti, G., \& Matteucci, F. 2011, A{\&}A,
  530, A33

\bibitem[{Vogt {et~al.}(1994)Vogt, Allen, Bigelow, Bresee, Brown, Cantrall,
  Conrad, Couture, Delaney, Epps, Hilyard, Hilyard, Horn, Jern, Kanto, Keane,
  Kibrick, Lewis, Osborne, Pardeilhan, Pfister, Ricketts, Robinson, Stover,
  Tucker, Ward, \& Wei}]{Vogt1994}
Vogt, S.~S., Allen, S.~L., Bigelow, B.~C., {et~al.} 1994, Soc. Photo-Optical
  Instrum. Eng. Conf. Ser., 2198, 362

\bibitem[{Vreeswijk {et~al.}(2013)Vreeswijk, Ledoux, Raassen, Smette, {De Cia},
  Wo?niak, Fox, Vestrand, \& Jakobsson}]{Vreeswijk2013}
Vreeswijk, P.~M., Ledoux, C., Raassen, A. J.~J., {et~al.} 2013, A{\&}A, 549,
  A22

\bibitem[{Watson {et~al.}(2006)Watson, Fynbo, Ledoux, Vreeswijk, Hjorth,
  Smette, Andersen, {Aoki, K.; Augusteijn}, Beardmore, Bersier, {Castro
  Cer{\'{o}}n}, D'Avanzo, Diaz-Fraile, Gorosabel, \& {Hirst, P.; Jakobsson, P.;
  Jensen, B. L.; Kawai, N.; Kosugi, G.; Laursen, P.; Levan, A.; Masegosa, J.;
  N{\"{a}}r{\"{a}}nen, J.; Page, K. L.; Pedersen, K.; Pozanenko, A.; Reeves, J.
  N.; Rumyantsev, V.; Shahbaz, T.; Sharapov, D.; Sollerman, J.; Starling, R. L.
  C.; Tanvi}}]{Watson2006}
Watson, D.~J., Fynbo, J. P.~U., Ledoux, C., {et~al.} 2006, ApJ, 652, 1011

\bibitem[{Weiner {et~al.}(2009)Weiner, Coil, Prochaska, Newman, Cooper, Bundy,
  Conselice, Dutton, Faber, Koo, Lotz, Rieke, \& Rubin}]{Weiner2009}
Weiner, B.~J., Coil, A.~L., Prochaska, J.~X., {et~al.} 2009, ApJ, 692, 187

\bibitem[{Werk {et~al.}(2016)Werk, Prochaska, Cantalupo, Fox, Oppenheimer,
  Tumlinson, Tripp, Lehner, \& Mcquinn}]{Werk2016}
Werk, J.~K., Prochaska, J.~X., Cantalupo, S., {et~al.} 2016, ApJ, 833, 54

\bibitem[{Werk {et~al.}(2012)Werk, Prochaska, Thom, Tumlinson, Tripp, O'Meara,
  \& Meiring}]{Werk2012}
Werk, J.~K., Prochaska, J.~X., Thom, C., {et~al.} 2012, ApJS, 198, 3

\bibitem[{White \& Frenk(1991)}]{White1991}
White, S. D.~M. \& Frenk, C.~S. 1991, ApJ, 379, 52

\bibitem[{White \& Rees(1978)}]{White1978}
White, S. D.~M. \& Rees, M.~J. 1978, MNRAS, 183, 341

\bibitem[{Wiseman {et~al.}(2017)Wiseman, Schady, Bolmer, Kr{\"{u}}hler, Yates,
  Greiner, \& Fynbo}]{Wiseman2017a}
Wiseman, P., Schady, P., Bolmer, J., {et~al.} 2017, A{\&}A, 599, A24

\bibitem[{Wolfe {et~al.}(2005)Wolfe, Gawiser, \& Prochaska}]{Wolfe2005}
Wolfe, A.~M., Gawiser, E., \& Prochaska, J.~X. 2005, ARA{\&}A, 43, 861

\bibitem[{Wolfe {et~al.}(1986)Wolfe, Turnshek, Smith, \& Cohen}]{Wolfe1986}
Wolfe, A.~M., Turnshek, D.~A., Smith, H.~E., \& Cohen, R.~D. 1986, ApJS, 61,
  249

\bibitem[{Woosley \& Bloom(2006)}]{Woosley2006a}
Woosley, S.~E. \& Bloom, J.~S. 2006, Annu. Rev. Astron. Astrophys, 44, 507

\end{thebibliography}
\renewcommand\thesection{\Alph{section}}
\onecolumn
\begin{appendix}
\setcounter{table}{0} \renewcommand{\thetable}{A.\arabic{table}}
\begin{longtab}

 \centering
 \small
 \newcolumntype{L}{>{\centering\arraybackslash}m{3cm}} 
 %\begin{tabular}{c c c c c c c c c}
 \begin{longtable}{c c c c c c c c c}
 \caption{\label{table:A}Ionic column densities of systems A, B, C, and D. Since high- and low-ionisation transitions tend to follow different profiles, we denote them differently: High-ionisation components are denoted with capital, Latin letters and low-ionisation components with Greek letters. Most do not trace the same component - exceptions are A$\alpha$=AB, B$\gamma \approx$BD, C$\alpha$=CB. $\ion{H}{I}$ typically follows the high-ions and as such is included in those components. }\\
 %\subcaption{ System A}

 \hline \hline
 \caption*{\label{table:all} System A}\\
 \hline \noalign{\smallskip}
 Component & & & & &  A$\alpha$& & &\\
 % \hline\noalign{\smallskip}
 $z$       & & & & &  3.362057& & & \\
 $v$ km s$^{-1}\tablefootmark{a}$ & & & & & +760&  &  & \\
 $b$ km s$^{-1}\tablefootmark{b}$ & &      &        &        &   5.0    &        &     &      \\
 \hline\noalign{\smallskip}
 Ion & Transitions Observed\tablefootmark{c}& \multicolumn{7}{c}{$\log(N)$ }\\
 $\ion{C}{II}$ & $\lambda$\textbf{1036}, $\lambda$\textbf{1334}&  & & & 13.32 $\pm$ 0.02 & &  & \\ 
 $\ion{C}{II}$* & $\lambda$1335& & & &   < 12.35&  & &\\	
 $\ion{O}{I}$ &$\lambda$1302 & & & &<12.05 & & & \\
 $\ion{Al}{II}$ & $\lambda$\textbf{1670}& & & & $<12.05$ & &  & \\ 
 $\ion{Si}{II}$ & $\lambda$\textbf{1260}& & & &   12.26 $\pm$ 0.03 & &  &\\          
 $\ion{Si}{II}$* & $\lambda$1265& & &   & < 12.25& & &\\       
 $\ion{Fe}{II}$ & $\lambda$1608 & & & & < 12.65 & & & \\  
 \hline	\noalign{\smallskip}
 Component &  & AG& AF& AE &AD & AC & AB & AA\\
	%\hline	\noalign{\smallskip}
 $z$ & & 3.359289 & 3.359745 & 3.360141 & 3.360861 & 3.361471 & 3.362057 & 3.362520 \\
 $v$ km s$^{-1}\tablefootmark{a}$ & &+570 & +601& +628&+678 &+720 &+760 &+792 \\
 $b$ km s$^{-1}\tablefootmark{b}$ & &  11 &   8  & 13  & 11  &  18 &   9   &   23  \\
 \hline	\noalign{\smallskip}
 Ion & Transitions observed\tablefootmark{c}& \multicolumn{7}{c}{$\log(N)$ }\\
 $\ion{H}{I}$&Lyman series \tablefootmark{d}&$\leq14.1$& $\leq15.5$&$\leq15.6$&$\leq15.8$&$\leq15.3$&$\leq16.2$&$\leq14.0$\\
 $\ion{C}{IV}$&$\lambda$\textbf{1548}, $\lambda$\textbf{1550} & 13.56 $\pm$0.01& >14.2 (S)& >14.7 (S)& > 14.7 (S)& 14.6 $\pm$0.02 & >14.4 (S)& 13.12 $\pm$ 0.07 \\
 $\ion{N}{V}$&$\lambda$1238, $\lambda$1242& & & & <12.0& & & \\
 $\ion{O}{VI}$&$\lambda$\textbf{1031}, $\lambda$\textbf{1037}&13.6 $\pm$0.7\tablefootmark{e} & 13.7 $\pm$0.3\tablefootmark{f}&- &- & 14.75 $\pm$0.01\tablefootmark{g} &- & -\\
 $\ion{Si}{IV}$ & $\lambda$\textbf{1393} ,$\lambda$\textbf{1402} & 12.92 $\pm$0.02 & > 13.7 (S) & 13.58 $\pm$0.01&13.72$\pm$0.01& 13.18 $\pm$0.02 & >13.7 (S) & 12.61 $\pm$0.05\\
 \hline\hline%\noalign{\smallskip}
 %\end{tabular}
 \end{longtable}
 %\subcaption{System B}
 %\begin{tabular}{c c c c c c c c}
 \begin{longtable}{c c c c c c c c}	 
 \caption*{\label{table:b} System B}\\
 \hline \noalign{\smallskip}
 Component& & &B$\delta$ & B$\gamma$&B$\beta$  &B$\alpha$ &\\	
 $z$      & & &3.350286& 3.351030&3.351123 & 3.351743 & \\
 $v$ km s$^{-1}\tablefootmark{a}$& & & -51  & 0& +6& +49 & \\
 $b$ km s$^{-1}\tablefootmark{b}$& & & 15&7 & 12 &    15  &\\
 \hline \noalign{\smallskip}
 Ion & Transitions Observed\tablefootmark{c}& \multicolumn{6}{c}{$\log(N)$ }\\
 $\ion{C}{II}$ & $\lambda$\textbf{1334}& &13.14$\pm$0.04&>14.5 (S)&>13.7 (S)  &13.78 $\pm$0.01 &\\
 $\ion{C}{II}$* & $\lambda$\textbf{1335}& &$12.65\pm0.1$&>14.5 (S)&>13.2 (S)  &13.33 $\pm$0.02&\\
 $\ion{N}{II}$ & $\lambda$\textbf{1083}& &13.03 $\pm$0.08 & 13.89$\pm$0.03& 13.29$\pm$0.07  &13.33$\pm$0.04 &\\
 $\ion{O}{I}$ &$\lambda$\textbf{1302}& &$<12.95$ & 13.65$\pm$0.02 & 12.93 $\pm0.14$ &$<13.1$ &\\
 $\ion{Al}{II}$&$\lambda$\textbf{1670}& &11.71 $\pm$0.05& >12.87 (S)& >11.79 & 12.22$\pm$0.02&\\
 $\ion{Si}{II}$&$\lambda$1190, $\lambda$1193, $\lambda$1260, & &12.23$\pm$0.06&13.82$\pm$0.02& <12.0 &12.85$\pm$0.08&\\
 & $\lambda$\textbf{1304}, $\lambda$\textbf{1526}& & & & & & \\
 $\ion{Si}{II}$*&$\lambda$\textbf{1194}, $\lambda$\textbf{1197}, $\lambda$\textbf{1264},&  & <11.0&12.87$\pm$0.03 &11.7$\pm$0.3 & 11.9 $\pm$0.1 &\\
 & $\lambda$\textbf{1309}, $\lambda$\textbf{1533}& & & & & & \\
 $\ion{Fe}{II}$& $\lambda$1608& & & <12.70 & & &\\
 \hline\noalign{\smallskip}
 Component& &BF& BE & BD & BC & BB & BA\\
	%\hline \noalign{\smallskip}
 $z$      & & 3.349792&3.350286& 3.351016&3.351349 &3.351630 & 3.351808 \\
 $v$ km s$^{-1}\tablefootmark{a}$& & -83& -51  & -1& 22&41 & 54 \\
 $b$ km s$^{-1}\tablefootmark{b}$& &10 &11&7 & 33 & 4&    8 \\
 \hline\noalign{\smallskip}
 Ion & Transitions Observed\tablefootmark{c}& \multicolumn{6}{c}{$\log (N)$ }\\
$\ion{H}{I}$&Lyman series \tablefootmark{d}&$\leq15.0$&$\leq15.0$ &$17.9 <N_{\ion{H}{I}}<18.35$ (S)&$\leq16.15$&$\leq16.0$ (S)\tablefootmark{h}&\\
 $\ion{C}{IV}$&$\lambda$\textbf{1548}, $\lambda$\textbf{1550} &13.20$\pm$0.01 &13.20$\pm$0.02&13.39$\pm$0.02&13.28$\pm$0.02&12.7$\pm$0.06&13.42$\pm$0.02\\
 $\ion{N}{V}$&$\lambda$1238, $\lambda$1242& & & <12.3& & & \\
 $\ion{Si}{IV}$ & $\lambda$\textbf{1393}, $\lambda$\textbf{1402}&12.60$\pm$0.02&12.89$\pm$0.01&13.25$\pm$0.01&13.16$\pm$0.02&12.28$\pm$0.05&12.99$\pm$0.02\\
 \hline \hline %\noalign{\smallskip}	
 %\end{tabular}
 \end{longtable}
 %\subcaption{System C}
 %\begin{tabular}{c c c c c}
 \begin{longtable}{c c c c c}
 \caption*{\label{table:c} System C}\\
 \hline \noalign{\smallskip}
 Component& &&C$\alpha$& \\
 $z$      & &&3.348545&\\
 $v$ km s$^{-1}\tablefootmark{a}$&& & -171&   \\
 $b$ km s$^{-1}\tablefootmark{b}$&&& 4 & \\
 \hline\noalign{\smallskip}
 Ion & Transitions Observed\tablefootmark{c}& &$\log(N)$&\\
 $\ion{C}{II}$ & $\lambda$\textbf{1334}&&13.50$\pm$0.02 &\\
 $\ion{C}{II}$* & $\lambda$\textbf{1335}&& 13.11$\pm$0.03 &\\
\caption*{\tablename{} (cont)}\\
 Ion & Transitions Observed\tablefootmark{c}& &$\log(N)$&\\
 $\ion{N}{II}$ & $\lambda$\textbf{1083}&&$<12.75$&  \\
 $\ion{O}{I}$ &$\lambda$\textbf{1302}& &<11.2 &  \\
  $\ion{Al}{II}$&$\lambda$\textbf{1670}&&11.66 $\pm$0.04&\\
 $\ion{Si}{II}$&$\lambda$\textbf{1190}, $\lambda$\textbf{1193}, $\lambda$1260, && 13.04$\pm$0.03&\\
 & $\lambda$\textbf{1304}, $\lambda$\textbf{1526}&  \\
 $\ion{Si}{II}$*&$\lambda$\textbf{1194}, $\lambda$\textbf{1197}, $\lambda$\textbf{1264},&&   11.7$\pm$0.1&\\ 
 $\ion{Fe}{II}$& $\lambda$1608&&<12.75 & \\
 \hline\noalign{\smallskip}
 Component&  & CC & CB & CA\\
 %\hline \noalign{\smallskip}
 $z$      & & 3.348234 & 3.348552 & 3.348649 \\
 $v$ km s$^{-1}\tablefootmark{a}$& & -193 & -171& -164 \\
 $b$ km s$^{-1}\tablefootmark{b}$& & 22&41 & 54 \\
 \hline\noalign{\smallskip}
 Ion & Transitions Observed\tablefootmark{c}& \multicolumn{3}{c}{$\log(N)$ }\\
 $\ion{H}{I}$&Lyman series \tablefootmark{d}&$<13.8$ & $15.53\pm0.05$\tablefootmark{i}&$<14.0$\\
 $\ion{C}{IV}$&$\lambda$\textbf{1548},  $\lambda$\textbf{1550} &13.2$\pm$0.2 &13.27$\pm$0.06&13.2$\pm$0.2\\
 $\ion{N}{V}$&$\lambda$1238, 1242& & $<13.05$&  \\
 $\ion{Si}{IV}$ & $\lambda$\textbf{1393}, $\lambda$\textbf{1402}&12.21$\pm$0.08&12.66$\pm$0.02&12.01$\pm$0.1\\
 \hline\hline
 \end{longtable}
	
 \begin{longtable}{c c c c c c}
 \caption*{\label{table:d} System D}\\
 \hline \noalign{\smallskip}
 Component&&DD&DC&DB&DA\\
 $z$     & &3.345680&3.346103& 3.346212&3.346816\\
 $v$ km s$^{-1}\tablefootmark{a}$&& -369& -339&-332 & -290\\
 $b$ km s$^{-1}\tablefootmark{b}$& & & & 4  & \\
 \hline\noalign{\smallskip}
 Ion &Transitions Observed\tablefootmark{c}& & &$\log(N)$& \\
 $\ion{H}{I}$ &Lyman series \tablefootmark{d}&$14.00\pm0.1$&$15.75\pm0.05$&$15.55\pm0.05$&$14.35\pm0.1$\\
 $\ion{C}{II}$ & $\lambda$\textbf{1036}, $\lambda$\textbf{1334}&&&12.66$\pm$0.05& \\
 $\ion{N}{II}$ & $\lambda$1083&&&$<12.15$& \\
 $\ion{N}{V}$ & $\lambda$1238, $\lambda$1242&&&$<11.85$& \\
$\ion{C}{IV}$ & $\lambda$1548, $\lambda$1550&&&$<12.05$&\\
 $\ion{O}{I}$ &$\lambda$1302&&&<12.5&\\
 $\ion{Al}{II}$&$\lambda$1670&&&$\leq11.2$&\\
 $\ion{Si}{II}$&$\lambda$1190, $\lambda$1260 &&&$\leq11.65$\tablefootmark{j}&\\
 $\ion{Si}{II}$&$\lambda$1309&&&$<12.4$\\
 $\ion{Si}{IV}$&$\lambda$1393, $\lambda$1402 &&&$<11.3$&\\
 $\ion{Fe}{II}$&$\lambda$1608 & &&$<12.5$&\\
\hline\hline
\end{longtable}
%\end{tabular}
 \tablefoot{(S): Measurement affected by saturation; (B): Measurement affected my blending; \tablefoottext{a}{Relative velocity compared to component B$\gamma$ at $z=3.351030$};  \tablefoottext{b}{$b$ is the broadening parameter for the component determined from the fit to the highest S/N, unsaturated and unblended lines};\tablefoottext{c}{Only transitions in bold have been used in the determining of column densities or upper limits};\tablefoottext{d}{see Section \ref{subsubsec:HI} and Figs \ref{fig:lyman_BCD},\ref{fig:lyman_A}};\tablefoottext{e,f,g}{$z=3.358461,3.359810,3.361463$ and $b=65,36,72$ \kms  respectively.}; \tablefoottext{h}{Combined upper limit for BA+ BB + BC};\tablefoottext{i}{$b=8.0$ \kms}; \tablefoottext{j}{The true nature of this measurement is discussed in Section \ref{subsec:D}}}
 %\end{table*}

\end{longtab}

\end{appendix}

\end{document}